\documentclass[a4paper,fleqn,usenatbib]{mnras}

\usepackage{newtxtext,newtxmath}

\usepackage[T1]{fontenc}
\usepackage{ae,aecompl}

\usepackage{graphicx}	
\usepackage{amsmath}	
\usepackage{amssymb}	

\newcommand{\kms}{\,km\,s$^{-1}$} 
\newcommand{\cmc}{\,cm$^{-3}$} 

\title[Isentropic instability in PDRs]{Isentropic thermal instability in atomic surface layers of photodissociation regions}

\author[ K.V. Krasnobaev and R. R. Tagirova]{K.V. Krasnobaev$^{1,2}$\thanks{Contact e-mail: kvk-kras@list.ru}
and R. R. Tagirova$^{1}$\thanks{rtaghirova@gmail.com}
\\
$^{1}$Space Research Institute of the Russian Academy of Sciences (IKI), Profsoyuznaya 84/32, Moscow 117997, Russia\\
$^{2}$Faculty of Mechanics and Mathematics, Lomonosov Moscow State University,  Leninskie Gory GSP-1, Moscow 119991, Russia
}

\date{Accepted XXX. Received YYY; in original form ZZZ}

\pubyear{2016}

\begin{document}
\label{firstpage}
\pagerange{\pageref{firstpage}--\pageref{lastpage}}
\maketitle

\begin{abstract}
We consider the evolution of an isentropic thermal instability in the atomic zone of a photodissociation region (PDR). In this zone,  gas heating and cooling are associated mainly with  photoelectric emission from dust grains and  fine-structure lines ([\ion{C}{ii}] 158, [\ion{O}{i}] 63, and [\ion{O}{i}]  146 {\micron}), respectively. The instability  criterion has a multi-parametric dependence on the conditions of the interstellar medium. We found that instability occurs when the intensity of the incident far-ultraviolet field $G_0$ and gas density $n$ are high. For example, we have  $3\times10^3<G_0<10^6$ and $4.5\times10^4<n<10^6$ {\cmc} at temperatures $360 <T<10^4$ K for typical carbon and oxygen abundances $\xi_{\rm C}=1.4\times10^{-4}$ and $\xi_{\rm O}=3.2\times10^{-4}$. The instability criterion depends on the relation between $\xi_{\rm C}$ and $\xi_{\rm O}$ abundances and line opacities. We also give examples of observed PDRs where instability could occur. For these PDRs, the characteristic perturbation growth time is $t_{\rm inst}\sim10^3$ -- $10^4$ yr  and the distance characterizing the formation of secondary waves is $L\sim10^{-3}$ -- $5\times10^{-2}$ pc. For objects that are older than $t_{\rm inst}$ and have sizes of the atomic zone larger than $L$, we expect that instability influences the PDR structure significantly. The presence of multiple shock waves, turbulent velocities of  several kilometers per second and inhomogeneities  with higher density and temperature  than the surrounding medium can characterize isentropic thermal instability in PDRs.
\end{abstract}




\section{Introduction}
 
A presentation of thermal instability is given by classical articles \citep{Parker1953, Zanstra1955, Field1965}, in which different types of instability are derived within linear theory.  Usually, in the study of the structure of the interstellar medium, the isobaric mode of thermal instability was considered \citep{BarKrasn1977, KaplanPikelner1979, OsterbrockFerland2006}. For example, the result of evolution of this mode was proposed to explain the observed two-phase structure (the co-existence of cold clouds and warm intercloud medium in pressure equilibrium) of the diffuse atomic interstellar medium  \citep{Field1969, Wolfire1995, Wolfire2003}. The criterion for the isobaric mode is stated in terms of the derivative of the generalized heat-loss function $Q$ at a constant pressure $p_0$ 
\begin{equation}
\frac{\partial Q}{\partial T} \Bigg|_{p_0} =\Bigg( \frac{\partial Q}{\partial T} - \frac{\rho}{T} \frac{\partial  Q}{\partial \rho} \Bigg) \Bigg|_{ \rho_0, T_0} > 0 ,
\label{eq:1}
\end{equation}
where  $Q=\Gamma-\Lambda$ is defined as the energy gain $\Gamma$ minus energy
loss $\Lambda$ (in erg g$^{-1}$ g$^{-1}$) in a static medium of density $\rho_0$ and temperature  $T_0$ (i.e. $Q(\rho_0,T_0)=0$). Condition \ref{eq:1},  in the limit of small  $Q$ corresponds to entropy perturbations. 

Significantly fewer articles are devoted to another type of thermal instability, the isentropic mode (also known as acoustic instability). This is due to the fact, that to satisfy the condition for this mode, special behaviour of the heat-loss function $Q$ is required (for more details, see Section \ref{sec:2.3}).  The criterion for the isentropic mode   is   stated in terms of the derivative  of   $Q$ at constant entropy $s_0$:
\begin{equation}
 \frac{\partial Q}{\partial T}\Bigg|_{s_0} =
 \Bigg( \frac{\partial  Q}{\partial T} + \frac{\rho}{(\gamma-1) T} \frac{\partial  Q}{\partial \rho} \Bigg) \Bigg|_{ \rho_0, T_0} > 0,
	\label{eq:2}
\end{equation}
where $\gamma$ is the adiabatic index.   In the limit of small  $Q$ condition \ref{eq:2} corresponds to nearly  adiabatic acoustic waves (i.e. adiabatic perturbations). 
 
For the interstellar medium, acoustic  instability was first studied in the article of \citet{Oppenheimer1977} for the molecular zone of photodissociation regions (PDRs). Further, this instability was discovered by \citet{Shchekinov1979} for the gas behind a radiating  shock wave. The problems of  non-linear evolution of isentropic perturbations  were considered by \citet{KrasnobaevTarev1987}. They found that non-linear steepening of a wave occurs due to the growth of perturbations and it is accompanied by formation of a shock wave. The effects of non-linear steepening of a wave in magnetized plasmas were explored by \citet{Nakariakov2000}.  Applying the Oppenheimer model, \citet{Krasnobaev1994} found that a sequence of self-sustained shock waves (also known as autowaves) is formed. \citet{Molevich2011}  investigated analytically and numerically the non-linear evolution and structure of plane autowaves in the atomic surface layer of a PDR. However, they considered only one case with density $n\sim10^3$ {\cmc} and incident far-ultraviolet flux $G_0=10^2$ and  did not take into account cooling in the oxygen  fine-structure lines, which becomes significant under these conditions  \citep{Wolfire1995}. Moreover, observations of PDRs indicate that $n$ and $G_0$ vary within a very wide range of parameters \citep{HollenbachTielens1999, Okada2013}, which will be considered below.

Thus,  the structure of our article is as follows. We present a model of energy balance in the atomic zone of a PDR. The model includes fine-structure emission in the carbon and oxygen lines; see Section  \ref{sec:2}. Based on this model, we define conditions when the steady-state $Q(\rho_0,T_0)=0$ satisfies  criterion \ref{eq:2}. We analyse wide ranges of the far-ultraviolet field $10<G_0<10^6$, gas densities $10<n<10^6$ {\cmc} and temperatures $10<T<10^4$ K, see Section \ref{sec:3}.  We use  the results of previous sections to identify astrophysical objects with parameters corresponding to adiabatic perturbations and we analyse  the ability for  instability  to occur  in them; see Section \ref{sec:4}.

\section{Energy balance}
\label{sec:2}

Photodissociation regions are regions where the energy balance and gas chemistry are determined mainly by far-ultraviolet radiation (FUV) in the range $6$ -- $13.6$ eV. For example,  a PDR is often formed at  the surface of a neutral molecular cloud, which is close to  young stars of O or B type. The general structure of a PDR has been studied in sufficient detail \citep{TielensHollenbach1985, Tielens2005} and can be described as follows. The medium around stars is ionized due to the radiation of photon energies larger than $13.6$ eV;  thereby  a region of ionized hydrogen (\ion{H}{ii}) is formed. We consider the structure of a PDR assuming that the \ion{H}{ii} region has reached pressure equilibrium with the surrounding medium. Radiation with energy $<13.6$ eV penetrates into the interstellar medium before the ionization front, dissociates molecular hydrogen H$_2$ in the Lyman and Werner bands ($11.2$ -- $13.6$ eV) and ionizes carbon.  A neutral zone of atomic hydrogen (\ion{H}{i}) is formed;  it is characterized by small impurities of heavy elements, such as (mainly) carbon ions  (\ion{C}{ii}) and oxygen atoms (\ion{O}{i}). When the distance from  stars increases and the FUV flux reduces, a \ion{C}{ii} transition into carbon monoxide (CO) occurs in the molecular cloud. At a greater distance, atomic oxygen transforms into molecular O$_2$. In this article, we will focus on the \ion{H}{i} zone in a PDR (it is located between the ionization and dissociation fronts).

Heating of atomic gas can occur through the following main processes: the photoelectric effect on large molecules and small dust grains; photopumping of H$_2$ molecules followed by collisional de-excitation of the resulting vibrationally excited species; neutral carbon photoionization. The last process is usually negligible compared with photoelectric emission. However, the FUV-pumped H$_2$ emission at high densities ($n>10^{4-5}$ {\cmc}) can be important and it has the same order as the photoelectric effect when the Lyman and Werner radiation fields are absorbed by H$_2$ lines rather than by dust  \citep{Burton1990}. A comparison of the H$_2$ line and dust absorption rates can be obtained by the steady-state H$_2$ formation-destruction equation, i.e. if we examine the ratio of the dissociation rate to the H$_2$ formation rate (or the atomic-to-molecular density ratio), which takes into account the attenuation of radiation. This ratio is expressed in the simplest approximation (\citealt{DraineBertoldi1996, HollenbachTielens1999}; for more details see \citealt{Sternberg2014}) as the ratio of the incident FUV flux $G_0$ (measured in units of $1.6\times10^{-3}$ \,erg\,cm$^{-2}$\,s$^{-1}$: \citealt{Habing1968}) to the density of hydrogen nuclei $n$.  The critical value of $G_0/n$ is approximately equal to $0.04$ cm$^3$; it corresponds to atomic and molecule column densities of $N(\ion{H}{i})=N(H_2)\sim10^{21}$ cm$^{-2}$ in the dissociation front, or visual extinction $A_{\rm V}\sim1$. If $G_0/n$ exceeds the critical value, then dust opacity becomes important.  Thus, when  $G_0/n < 0.04$ cm$^3$  ($A_{\rm V}<1$), gas heating of the pumping H$_2$ is significant and, conversely, heating is unimportant for $G_0/n > 0.04$ cm$^3$ (the \ion{H}{i}/H$_2$ transition zone corresponds to $A_{\rm V}\sim1$--$2$). For typical PDRs we have the average value $G_0/n\sim0.1$--$1$ cm$^3$ \citep{Tielens2005}. Hereafter we will consider $G_0/n> 0.04$ cm$^3$.

According to the concepts of PDR structure, the energy balance of the \ion{H}{i} zone is  determined mainly by photoelectric heating from dust grains and gas cooling through infrared fine-structure lines of atoms and ions. In the next subsections, we consider the physical processes in detail.

\subsection{Heating}

Photoelectric emission from dust grains and polycyclic aromatic hydrocarbon molecules (PAH) dominates heating in the atomic zone of PDRs. Photoelectric heating from interstellar grains (for brevity, the PAH will be called grains) was first described by \citet{Spitzer1948}. This description  was improved by  \citet{TielensHollenbach1985, BakesTielens1994, Wolfire2003, WeingartnerDraine2001b}. We use the modification of the heating $\Gamma_{\rm pe}$ proposed by \citet{WeingartnerDraine2001b}, which takes dust grain-size distributions into account.  We also consider the energy loss $\Lambda_{\rm pe}$ in the gas due to the accretion of charged particles on to the grains (it is significant for high temperature $T>10^3$ K).  Heating and cooling are reproduced by the following functions

\[
\begin{split}
&\Gamma_{\rm pe}=10^{-26} \, \, {\rm \,ergs\,s^{-1}} \\
& \times \frac{G_0}{m_{\rm H}}\,
\frac{C_0+C_1T^{C_4} }{1+C_2(G_0 \sqrt{T}/n_e)^{C_5} (1+C_3 (G_0 \sqrt{T} /n_e)^{C_6} )} \, ,
\end{split}
\]
\[
\begin{split}
&\Lambda_{\rm pe}=10^{-28} \, \, {\rm \,ergs\,cm^3\,s^{-1}}  \frac{n_e}
{m_{\rm H}} \,T^{(D_0+D_1/\chi)} \\
& \times  \exp(D_2+D_3 \chi - D_4 \chi^2 )  \, \, \, \, \, \,  {\rm for} \, \, \, \,  \chi=\ln(G_0 \sqrt{T}/n_e) \, ,
\end{split}
\]
where $n_e$ denotes the number density of electrons and $m_{\rm H}$ is the mass of the hydrogen atom. Almost all carbon near the surface of PDRs is ionized, hence $n_e=\xi_{\rm C}\,n$, where $n$ is the number density of hydrogen and $\xi_{\rm C}$ is the carbon abundance in the gas.  Coefficients $C_0,...,C_6$ and $D_0,...,D_4$ are given in \citet{WeingartnerDraine2001b} and depend on  the dust properties (grain size, composition) and a radiation field spectrum. 

According to \citet{WeingartnerDraine2001a}, grain size distributions are consistent with the observed extinction of starlight, which varies depending on the environment through which light travels. Extinction variations can be parameterized by the ratio of visual extinction to reddening $R_{\rm V}=A_{\rm V}/E_{\rm B-V}$  \citep{CardelliClaytonMathis1989}. A diffuse interstellar medium with density $n\leqslant10^2$ {\cmc} corresponds to $R_{\rm V}\sim3.1$; higher values $R_{\rm V}\sim5$--$6$ are observed for dense clouds $n>10^4$ {\cmc} and intermediate-density regions correlate with $R_{\rm V}\sim4$. Moreover, \citet{WeingartnerDraine2001a} showed that grain size distributions reproduce the observed extinction better if the contribution of very small carbonaceous grains is considered. They constructed the size distributions for various combinations of $R_{\rm V}$ and $b_{\rm C}$, where  $b_{\rm C}$ is the C abundance (per H nucleus) in very small grains (radius $\leqslant 100$ \AA). \citet{LiDraine2001} found that the emission observed from dust in the diffuse interstellar medium and the corresponding extinction curve agree better when $b_{\rm C}$ reaches the maximum value from all possible variations at the given $R_{\rm V}$ (i.e. $b_{\rm C}=6\times10^{-5}$ at  $R_{\rm V}=3.1$). \citet{WeingartnerDraine2001a} suggest that this assumption also holds in denser regions, therefore the largest allowed values $b_{\rm C}=4\times10^{-5}$ and $b_{\rm C}=3\times10^{-5}$ can be used for $R_{\rm V}=4$ and 5.5, respectively. Hereafter, for simplicity these combinations will be considered. However, as the application of these results to  observations, we provide an example with $b_{\rm C}=0$ and $R_{\rm V}=5.5$ (see Section \ref{sec:4}, Carina N).

In addition, we make the following assumptions. First, the grain size distributions are constructed so as to minimize the influence of  carbon and silicate inclusions (case A by \citealt{WeingartnerDraine2001a}). Secondly, we adopt a blackbody radiation field with colour temperature $T_{\rm_c}=3\times10^4$ K.

As a result, the total photoelectric heating is represented as
\[\Gamma(n,T,G_0,\xi_{\rm C},R_{\rm V},b_{\rm C})=\Gamma_{\rm pe}-\Lambda_{\rm pe} \]
The function $\Gamma$ is mainly dependent on the gas temperature $T$ and grain charge parameter $G_0 \sqrt{T}/n_e$, which characterizes the ratio of ionization and  recombination rates of grains. An increase in $G_0 \sqrt{T}/n_e$ leads to a higher grains charge and therefore heating efficiency $\Gamma/G_0$  decreases \citep{BakesTielens1994}. Properties of the gas-dust medium show less heating efficiency for dense regions characterized by $R_{\rm V}=5.5$ ($b_{\rm C}=3\times10^{-5}$) than for  diffuse regions with $R_{\rm V}=3.1$ ($b_{\rm C}=6\times10^{-5}$) \citep{WeingartnerDraine2001b}.

\subsection{Cooling}
\label{subsec:2.2}

The  atomic gas of PDRs is cooled predominantly through the fine-structure excitation of ions and atoms by atomic hydrogen impact. The largest contribution to the  gas cooling comes from the  [\ion{C}{ii}] 158, [\ion{O}{i}] 63 and [\ion{O}{i}]  146 {\micron} lines  \citep{TielensHollenbach1985, Hollenbach1991, Tielens2005}. The radiative cooling rate due to the transition from upper  level 2 to lower level 1 of some species is given by
\begin{equation*}
\begin{split}
& \Lambda_{21}=\xi 
\, E_{21} \, A_{21} \, \beta(\tau_{21})  \\ 
& \times  \frac{1 }{m_{\rm H}\left[1+{g_1}/{g_2} \, \exp({E_{21}}/{k_{\rm B} T}) (1+ \beta(\tau_{21}) \, {n_{\rm_{cr}}}/{n})\right]} \, ,  
\end{split}
\end{equation*} 
where $E_{21}$ is the energy difference between two levels, $A_{21}$ is the spontaneous transition probability, $g_2$ and $g_1$ are the statistical weights of two levels, $k_{\rm B}$ is the Boltzmann constant and $\xi$ is the abundance ($\xi_{\rm C}$ for carbon and $\xi_{\rm O}$ for oxygen). The critical density for de-excitation processes is $n_{\rm_{cr}}=A_{21}/\gamma_{21}$, i.e. roughly the density above which the levels thermalize collisionally. Here, $\gamma_{21}$ is the collisional de-excitation rate coefficient for atomic hydrogen collisions (Table~\ref{tab:1}). Parameter $\tau_{21}$ is the optical depth averaged over the line and  $\beta(\tau_{21})$ is an escape probability at optical depth $\tau_{21}$ of the line. In the limit of small optical depth, $\beta\sim0.5$ (in a semi-infinite slab); at large optical depth, $\beta\sim1/\tau_{21}$ \citep{Jong1980}.

\begin{table}
	\centering
	\caption{Parameters of the cooling}
	\label{tab:1}
	\begin{tabular}{lcccc} 
		\hline
		Spacies & ${\lambda_{21}}^a$ &  $E_{21}$  & $A_{21}$ & $\gamma_{21}$ \\ 
		& ({\micron})& (K) & (s$^{-1})$ &(cm$^3$\,s$^{-1}$)\\
		\hline
		\ion{C}{ii} & 158 & 92 & $2.4\times10^{-6}$&$ 8.12\times10^{-10} T^{0.02}$\\
		\ion{O}{i} & 63 & 228 & $8.95\times 10^{-5}$&$ 4.2\times10^{-12} T^{0.67}$\\
		\ion{O}{i} & 146 & 98 & $1.7\times 10^{-5}$&$1.45\times10^{-11} T^{0.44}$\\
		\hline   
	\end{tabular}
\begin{tabular}{l}
Note. 
$^a$ The wavelenght of $2\rightarrow1$ transition			
\end{tabular}
\end{table}

Next, we will estimate approximately the relations between optical depths of lines. Let $N_{\tau{21}}$ be the column density of hydrogen nuclei required for unit optical depth in the level $2\rightarrow$ level $1$ transition under the assumption that all atoms or ions of the corresponding element are in the lower level (an expression for $N_{\tau{21}}$ can be found in \,\citealt{TielensHollenbach1985}), i.e. $\sigma_{21} \xi  N_{\tau{21}}$=1, where $\sigma_{21}$ is the cross-section absorption for the level $2\rightarrow$ level $1$ transition. For any $\tau_{21}$, we introduce $\tau_{21}= N/N_{\tau{21}}$, where $N$ is the column density of hydrogen nuclei.
For [\ion{C}{ii}] 158, [\ion{O}{i}] 63, and [\ion{O}{i}]  146 {\micron}  lines, we denote the column densities $N_{\tau{21}}$ as  $N_{\rm \tau{C}}$, $N_{\rm \tau{O63}}$ and $N_{\rm \tau{O146}}$, respectively, and consequently we have the optical depths  $\tau_{21}$ as $\tau_{\rm C}=N/N_{\tau{\rm C}}$, $\tau_{\rm O63}=N/N_{\tau{\rm O63}}$, and $\tau_{\rm O146}=N/N_{\rm \tau{O146}}$, respectively. Therefore,  $\tau_{\rm O63}/\tau_{\rm C}=N_{\rm \tau{C}}/N_{\rm \tau{O63}}=0.72\times \xi_{\rm O}/\xi_{\rm C}$  and $\tau_{\rm O146}/\tau_{\rm C}=N_{\rm \tau{C}}/N_{\rm \tau{O146}}=0.92\times \xi_{\rm O}/\xi_{\rm C}$. 
In the  \ion{H}{i}  zone of PDRs we usually have $\tau_{\rm C}<1$  \citep{Tielens2005}.

Data of the carbon C and oxygen O abundances varies for different photodissociation regions. According to observations, the ratio  $\xi_{\rm O}/\xi_{\rm C}$ is approximately equal to two. For example, values  of $\xi_{\rm C}$ are assumed to be $\xi_{\rm C}=1.4\times 10^{-4}$ \citep{Cardelli1996} or $\xi_{\rm C}=1.6\times 10^{-4}$ \citep{Sofia2004}, while $\xi_{\rm O}=3.2\times 10^{-4}$ \citep{Meyer1998}. However, there are also higher estimates for the abundances: for instance, in the Orion Bar:  $\xi_{\rm C}=3\times 10^{-4}$ and $\xi_{\rm O}=(4$ -- $5)\times 10^{-4}$  \citep{Wolfire1995, Shaw2009}.

Total energy losses in the lines considered are represented by
\[
 \Lambda(n,T, \xi_{\rm C}, \xi_{\rm O} ,\tau_{\rm C})= \Lambda_{\rm CII158}+ \Lambda_{\rm OI63}+\Lambda_{\rm OI146}
 \]
 Radiative cooling has its largest value when the optical depth is small, i.e. $\tau_{21} \rightarrow 0$  ($\beta \sim0.5$). This may occur near the boundary of the PDR and \ion{H}{ii}  region.
 
Thus, we have shown that generalized heat-loss function $Q=\Gamma-\Lambda$ depends on the parameters of the gas-dust medium and the radiation passing through it (i.e. $n, T, G_0, R_{\rm V}, b_{\rm C}$); also, $Q$ depends on the cooling line opacity ($\tau_{\rm C}$) and the abundances of heavy elements ($\xi_{\rm C}, \xi_{\rm O}$).

\subsection{Isentropic criterion}
\label{sec:2.3}

\begin{figure}
\begin{minipage}{0.49\columnwidth}
\center{\includegraphics[width=\columnwidth]{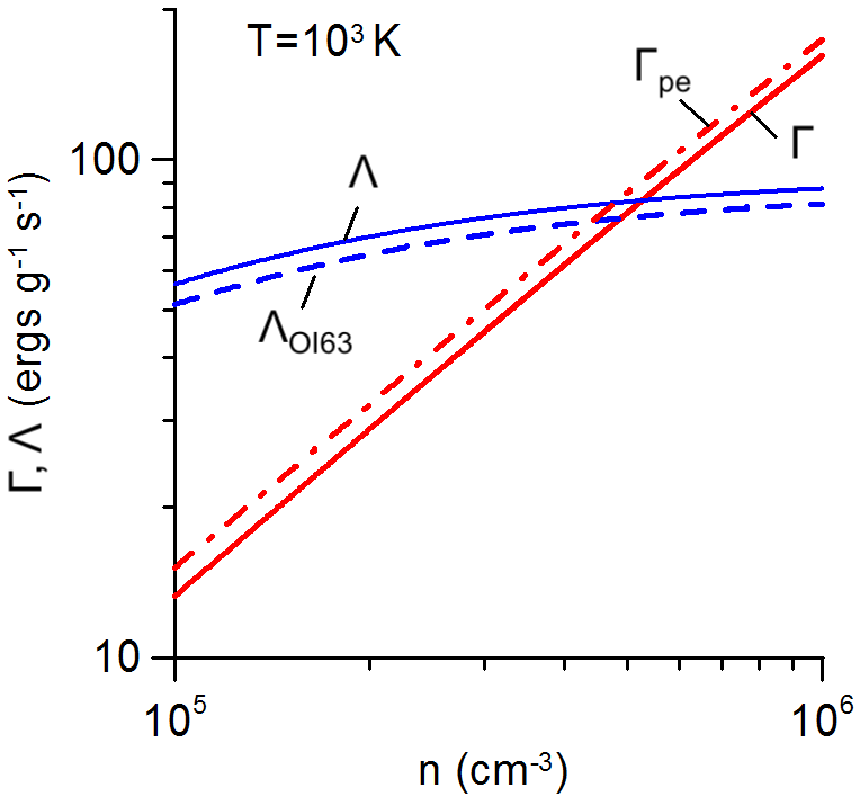}  \\ (a)}
\end{minipage} 
\begin{minipage}{0.45\columnwidth}
\center{\includegraphics[width=\columnwidth]{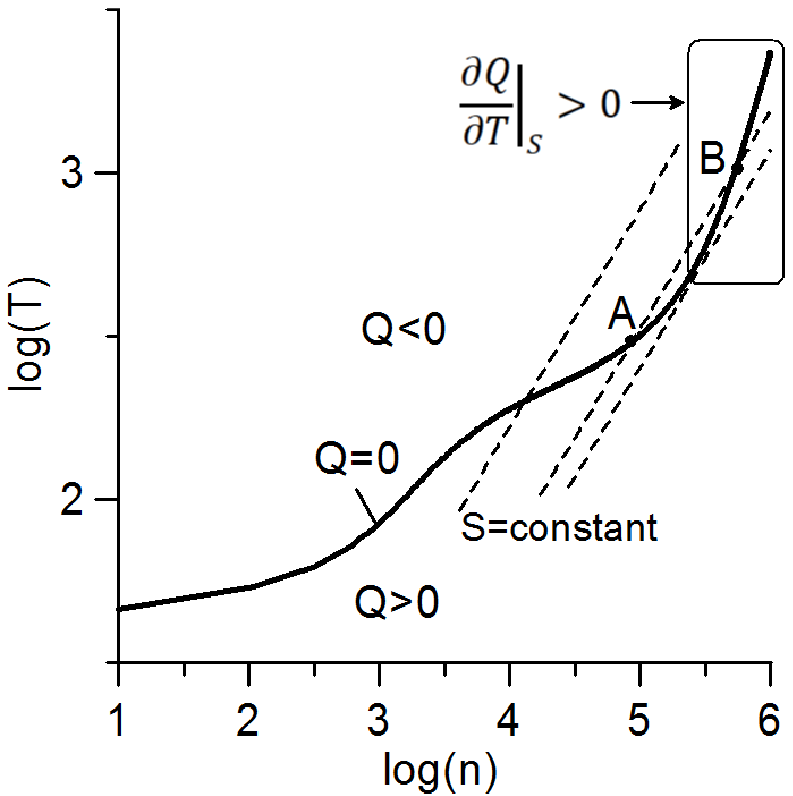}  \\ (b)}
\end{minipage}
    \caption{Behaviour of heating and cooling functions ($G_0=10^5$, $R_{\rm V}=5.5$, $b_{\rm C}=3\times10^{-5}$, $\xi_{\rm C}=1.4\times10^{-4}$, $\xi_{\rm O}=3.2\times10^{-4}$ and $\beta=0.5$): (a)  total heating $\Gamma$ and cooling $\Lambda$ rates at $T=10^3$ K (where $\Gamma_{\rm pe}$ and  $\Lambda_{\rm OI63}$ are dominant processes); (b) contour of thermal balance, $Q(n, T) = 0$ (solid curve),  with the locus for constant entropy $s$ (dashed line). The locus only inside a square (top right) is thermally unstable by the isentropic criterion.}
\label{fig:1}
\end{figure}

For interstellar gas, acoustic instability was  first demonstrated by  \citet{Oppenheimer1977}.
He noted that this instability can be understood as  the preferential heating of compressed regions of sound wave. It  happens if a heating rate (in ergs cm$^{-3}$ s$^{-1}$) is an increasing function of  $n$ or $T$ under conditions where a cooling rate is relatively insensitive to $n$ or $T$ (see Fig. \ref{fig:1}(a) for our model of energy balance). Oppenheimer found such  conditions in the molecular regions of PDRs, where the molecular transitions governing the cooling of the gas are thermalized (this occurs at high density)  and strong heat sources are present. Here, the heating rate  usually varies at least as rapidly as $n$ and the cooling rate is almost independent of density.  Notice that at  high density the sign of the derivative $\partial Q/\partial \rho$ determines the sign of the isentropic  criterion \ref{eq:2}. 

We shall verify that similar conditions are satisfied for the atomic zone of  PDRs at high density. Indeed, we can see that photoelectric heating is an increasing function of  density $n$  \citep{BakesTielens1994}  
and the cooling rate  depends weakly on $n$ when $n>n_{\rm cr}$. The justification for the behaviour of the cooling rate can be as follows. The line [\ion{O}{i}] 63 $\micron$ becomes an important component of the total cooling rate  with the increase of  density $n$ and FUV field $G_0$ (where $G_0$ influences the steady-state temperature $T_0$) and it becomes  dominant at high $n$  and  $G_0$ \citep{TielensHollenbach1985}.   For the  cooling line   [\ion{O}{i}] 63 $\micron$,  we have  the value $n_{\rm cr} \sim 10^5$ \cmc (where $n_{\rm_{cr}}=A_{21}/\gamma_{21}$, see Table~\ref{tab:1}). At $n > n_{\rm  cr}$, the total  cooling rate depends  weakly on $n$.  This behaviour  of the rates  is shown in Fig. \ref{fig:1}(a).

Thus,  by analogy with  \citet{Oppenheimer1977}, we assume that the isentropic type can arise in the dense atomic zone of photodissociation regions. However it is quite a rough estimate. 
Exact knowledge of the conditions under which the  isentropic mode will grow can be obtained through a direct application of the corresponding criterion, i.e. checking the positivity of the derivative $(\partial Q/\partial T)|_{s_0}>0$. 

The locus of the heat-loss function $Q$ satisfying this  criterion is shown in  Fig. \ref{fig:1}(b). Assume that we have a static, homogeneous gas in thermal equilibrium at some $n$ and $T$. We have constructed  Fig. \ref{fig:1}(b) for typical parameters causing acoustic instability. The region above the curve of thermal balance corresponds to $Q<0$, because cooling exceeds heating if the temperature exceeds the equilibrium value for a given density. Conversely, the region below the curve corresponds to $Q>0$. We consider a small inhomogeneity embedded in this medium and perturb it away from the equilibrium curve along the locus $s\propto \ln(p/\rho^\gamma)={\rm constant}$ (where $p/\rho^\gamma \propto T/n^{\gamma-1}$).  Let the inhomogeneity exists at point $A$; we displace it slightly to lower (higher) temperatures and lower (higher)  densities along the locus $s = constant$. According to the diagram, the inhomogeneity  enters a region where $Q>0$ ($Q<0$), i.e. where the heating exceeds the cooling (or vice versa). Thus the inhomogeneity  must heat up (cool down)  again and re-expand back toward the point $A$. All gas located  in the  region $A$ is  thermally stable.  

Now let us consider the case if the inhomogeneity  exists in the square region in Fig. \ref{fig:1}(b), e.g. at the  point B. If we take a piece of such a medium and displace it toward
lower (higher)  temperatures and lower (higher)  densities, it will now enter a region where $Q<0$ ($Q>0$), i.e. where  cooling exceeds  heating (or vice versa). Thus, maintaining the same entropy as its surroundings, such a medium would get cooler (hotter) and more rarefied (denser),  until it makes a transition to the thermally stable state,  e.g. to the  region $A$ for $Q<0$, or until the heating and compression are stopped for some reason in the case $Q>0$. Gas placed in region $B$ is isentropic thermally unstable. A medium placed in the unstable region would therefore co-exist in two states,  cold rarefied gas and warm dense gas, at a common entropy $s$.   Investigations of acoustic perturbations  \citep{KrasnobaevTarev1987, Molevich2011} and also our calculations (see below) confirm these features. 

Notice that the heat-loss function $Q$ depends not only on the variables $n$ and $T$ but also on the set of parameters that define the conditions in the interstellar medium (i.e. the gas-dust properties and the radiation passing through the medium represented by $R_{\rm V}$, $b_{\rm C}$ and $G_0$,  
the cooling-line opacity represented by $\tau_{\rm C}$, and the abundances of heavy elements represented by $\xi_{\rm C}$ and $\xi_{\rm O}$). Therefore, to find the conditions  for isentropic instability growth, we calculate $(\partial Q/\partial T)|_{s_0}$ and find the conditions for its positivity (Section \ref{sec:3}). 

\section{PDR parameters causing instability}
\label{sec:3}

To study the instability evolution of travelling waves, we start with consideration of its general features. Thus in Section \ref{sec:unstable} we present a theoretical description of isentropic thermal instability, followed by a numerical simulation. As the PDR characteristics vary over a very wide range, in Section \ref{sec: detect}  we provide a multivariable analysis to show that the instability criterion \ref{eq:2}  is satisfied.

\subsection{Evolution of unstable perturbations}
\label{sec:unstable}

To describe the gas motion in the atomic zone of a PDR, we consider the system of gas dynamics equations 
\begin{equation*}
\begin{split}
&\frac{d \rho}{dt}+\rho \, \mathrm{div} \, \textbf{v}=0 \, \\
& \frac{d \textbf{v}}{dt} +\frac{1}{\rho} \mathrm{grad} \, p=0 \, \\
&\frac{d}{dt} \Big(\frac{p}{(\gamma-1)\rho} \Big)+\frac{p}{\rho} \, \mathrm{div} \, \textbf{v}=Q 
\end{split}
\end{equation*}
Here $\rho=n m_{\rm H}$, $p=\rho R T$, $t$ and ${\bf v}$ are the mass density, pressure, time and gas velocity, $R=k_{\rm B}/m_{\rm H}$ is the universal gas constant and $\gamma=5/3$ is the adiabatic index.   We consider one-dimensional plane motion with velocity $u$ along the $x$ coordinate.

The steady state  is characterized by $\rho=\rho_0$ and $T=T_0$ at $u=0$  such that $\Gamma(\rho_0,T_0)=\Lambda(\rho_0,T_0)=\Lambda_0$ and $Q(\rho_0,T_0)=0$.
We assume that the characteristic parameters of gas motion are the density $\rho_0$,  temperature $T_0$, isothermal sound speed $u_0=\sqrt{RT_0}$, time of cooling  $t_0=R T_0/\Lambda_0$ and length-scale $l_0=u_0 t_0$.

{We study  the short-wavelength regime of the wave mode of thermal instability found by \citet{Field1965}. In this case, the wave mode satisfies the isentropic criterion \ref{eq:2} and  its growth rate is given by the expression
\begin{equation*}
\omega=\frac{(\gamma-1)^2}{2\gamma R}\Big(\frac{\partial Q}{\partial T}+\frac{\rho}{(\gamma-1)T} \frac{\partial Q}{\partial T}\Big) \Big|_{\rho_0, T_0}\\
\end{equation*}
which is also similar to equation 1.8 in  \citet{KrasnobaevTarev1987} (where they use  $Q$  per unit volume and time, which  differs from our notation). The characteristic time of perturbation growth for the isentropic  type is $t_{\rm inst}=1/\omega$. 

The small-wavelength limit  is satisfied  when the time $t_{\rm inst}$ exceeds the sound-crossing time $t_{\rm s}=\lambda/\sqrt{\gamma} u_0$ $ \sim t_0 \lambda/l_0$  \citep{Vazquez2003}, where $\lambda$ is  a wavelength. For typical parameters of PDRs, cooling time $t_0 < 10^2$ yr and hence  $t_{\rm s}  < 10^2 \lambda/l_0$ yr, whereas  usually the time of perturbation growth $t_{\rm inst} > 10^2$ yr. Therefore, to satisfy the regime $t_{\rm inst} > t_{\rm s}$ we  assume, for  simplicity, that the wavelength $\lambda$  is of the same order as the characteristic length-scale $l_0$.

The condition  $t_{\rm inst} > t_{\rm s}$  permits us  to use the weak non-linear theory of \citet{KrasnobaevTarev1987}. This theory   allows us to study  the propagation of non-linear stationary waves of finite amplitude and verify  the simulation results.
}

The influence of  dissipative processes, e.g. thermal conductivity, on $t_{\rm inst}$ is seen in the existence of an upper limit on the wavenumber, above which the growth of perturbations is inhibited.
In the general case, the damping of perturbations in the short-wavelengths limit  follows from the theory of travelling waves in a thermal conducting medium, which was  investigated by \citet{Landau1987}. 
 Applied to thermal instability, the damping effect  was obtained  by \citet{Field1965}. The influence of thermal conductivity in PDRs is discussed in Section \ref{sec:4}.

\begin{figure*}
\includegraphics[width=2\columnwidth]{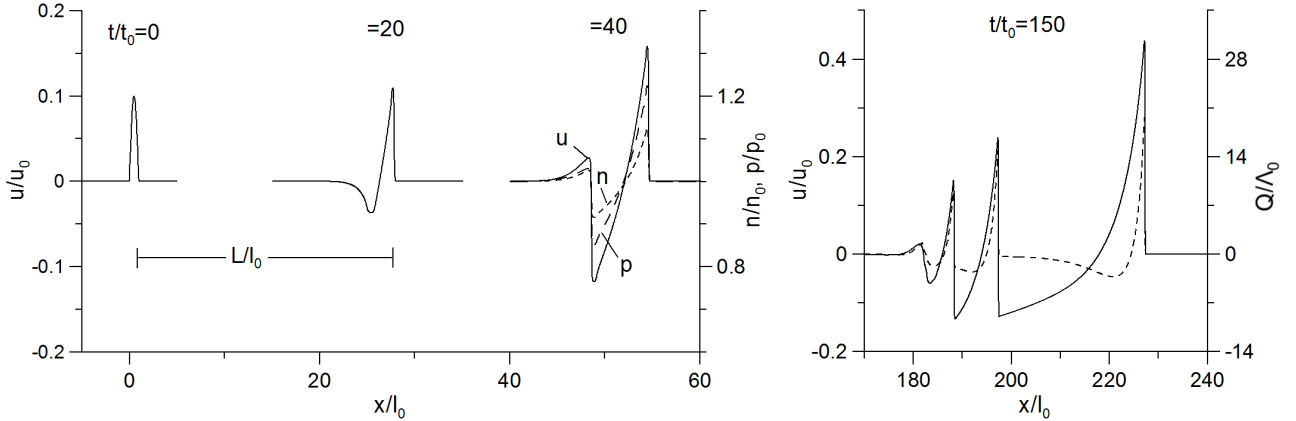}  
    \caption{Evolution of velocity perturbations $u$ for $n_0=5\times10^5$ {\cmc}, $T_0=943$ K, $G_0=10^5$, $R_{\rm V}=5.5$, $b_{\rm C}=3\times10^{-5}$, $\beta(\tau_{21})=0.5$, $\xi_{\rm C}=1.4\times10^{-4}$ and $\xi_{\rm O}=3.2\times10^{-4}$. The characteristic time of perturbation growth is $t_{\rm inst}=344$ yr,  $t_{\rm inst}/t_0=20$. The distributions of density $n$ and pressure $p$ at $t/t_0=40$ are shown, as is the generalized heat-loss function $Q$  (dashed curve) at $t/t_0=150$ ($\Lambda_0=1.47\times10^2$ erg\,g$^{-1}$\,s$^{-1}$).}
\label{fig:2}
\end{figure*}

Studies of the isentropic mode  \citep{KrasnobaevTarev1987, Krasnobaev1994, Molevich2011} show that the growth of initially small perturbations at the non-linear  evolution stage is accompanied by the formation of a sequence of self-sustained shock waves (autowaves). \citet{Krasnobaev2016} found numerically  that the waves reach saturation and hence have a maximum amplitude that is determined by the heat-loss function $Q$ and  depends weakly on the parameters of the initial perturbations (wavelength $\lambda$ and amplitude $a$). We assume that $\lambda=2 l_0$. Fig. \ref{fig:2} gives an example of perturbation evolution that begins with a  single pulse described by $u/u_0=a \sin(\pi x/\lambda)$ at $a=0.1$, $n/n_0=1$ and $p/p_0=1$ for $0<x<\lambda$. The wave evolution is calculated by the total variation diminishing Lax-Friedrichs scheme.

Fig.  \ref{fig:2} shows that, for the time  about $t_ {\rm inst}/t_0\sim20$, the velocity perturbation grows (also,  perturbations of $n$ and $p$ increase, which we can see at $t/t_0=40$) and then a shock wave forms. The gas state behind the initial perturbation is not steady and therefore a secondary wave arises. Consequently a sequence of shock waves is generated, which is shown at $t/t_0=150$. The function $Q$ 
in Fig.  \ref{fig:2} shows typical properties of isentropic oscillations. Thus perturbations are subject to a slight heating during the compression phase, which tends to increase the amplitude of the wave. 

We consider the distance $L$ between the source of the initial perturbation and the primary wave when the secondary wave begins to form (see Fig.  \ref{fig:2} at $t/t_0=20$).  We can estimate $L$ by the expression
 \begin{equation*}
 L \sim a_0 \, t_{\rm inst}  \, \, \, \,  \, \, \textrm{or} \, \, \, \,\,   L/l_0 \sim \sqrt{\gamma} \, t_{\rm inst}/t_0  
\end{equation*}
{where $a_0=\sqrt{\gamma R T_0}$}.
Notice that  the distance between the primary and secondary waves will increase with time, due to the difference between their velocities.

\subsection{Detection of parameters causing instability}
\label{sec: detect}

We consider the following parameters causing  instability: $n_0, T_0, G_0, R_{\rm V}, b_{\rm C}, \tau_{\rm C}, \xi_{\rm C}$, and $\xi_{\rm O}$, for which the heat-loss function $Q$ satisfies criterion \ref{eq:2}. The density $n$,  temperature $T$ and  FUV field $G_0$ of PDRs vary over wide ranges \citep{Tielens2005}:  
 \begin{equation}
10<n<10^6 \textrm{{\cmc}} \, , \, \, \,  10<T<10^4 \, \textrm{K} \, , \, \, \,  10<G_0<10^6 \,. 
 	\label{eq:3}
\end{equation}

We want to find the range of parameters causing instability for intervals \ref{eq:3} and  the values of $R_{\rm V}, b_{\rm C}, \tau_{\rm C}, \xi_{\rm C}$, and $\xi_{\rm O}$ considered in Section \ref{sec:2}. First, we consider variations  of $R_{\rm V}$ and $b_{\rm C}$, which characterize the dust properties for typical abundances of carbon $\xi_{\rm C}=1.4\times10^{-4}$ and oxygen $\xi_{\rm O}=3.2\times10^{-4}$. We also assume  small optical depths for the cooling lines. Secondly, we investigate  the influence of optical depths on the range of parameters causing instability. We vary $\tau_{\rm C}$ from 0 to 1, where $\tau_{\rm C}\sim0$ corresponds to the position of matter near the PDR surface, while  $\tau_{\rm C}=1$ corresponds to a position further into  the PDR. Third, we study the contribution of carbon C and oxygen O to the variations of parameters causing instability.

\subsubsection{$R_{\rm V}$ variations}
\label{sec:Rv}
As discussed in Section \ref{sec:2}, in diffuse regions the combination of $R_{\rm V}$ and $b_{\rm C}$ has the best agreement with observations of dust grain-size distributions when  $b_{\rm C}$ attains its largest allowed values. Therefore, we consider three typical combinations: $R_{\rm V}=3.1$, $b_{\rm C}=6\times10^{-5}$ -- diffuse interstellar medium; $R_{\rm V}=5.5$, $b_{\rm C}=3\times10^{-5}$ -- dense clouds; $R_{\rm V}=4$, $b_{\rm C}=4\times10^{-5}$ -- intermediate-density regions.

\begin{figure*}
\includegraphics[width=2\columnwidth]{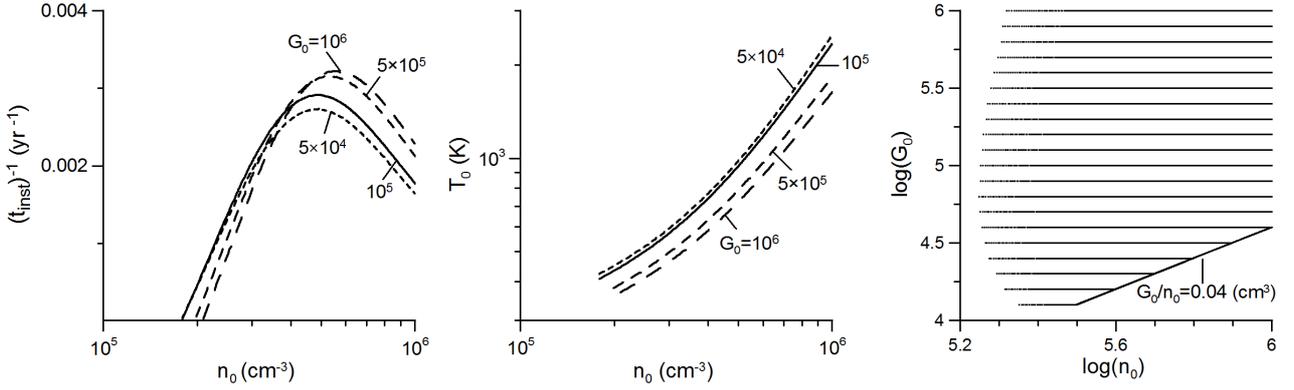}   
    \caption{ Examples where the isentropic instability criterion are satisfied ($R_{\rm V}=5.5$, $b_{\rm C}=3\times10^{-5}$, $\tau_{\rm C}\sim0$, $\xi_{\rm C}=1.4\times10^{-4}$ and $\xi_{\rm O}=3.2\times10^{-4}$). 
 {(a) Growth rate $1/t_{\rm inst}$},  (b) temperature $T_0$ and (c) FUV flux $G_0$  (we use the condition $G_0/n_0>0.04$ cm$^3$, for which photoelectric heating dominates in a dense gas).}
\label{fig:3}
\end{figure*}

Criterion \ref{eq:2}  in intervals \ref{eq:3} for $\xi_{\rm C}=1.4\times10^{-4}$ and  $\xi_{\rm O}=3.2\times10^{-4}$  shows that instability appears in dense regions with $10^5 \lesssim n_0<10^6$ {\cmc}. Such dense gas usually corresponds to high values of the ratios of visual extinction to reddening, for example $R_{\rm V}=5.5$. Smaller values, $R_{\rm V}=3.1$ and 4,  are characterized by smaller density, $n_0 \lesssim 10^4$ {\cmc}, while instability can occur only when $n_0 \gtrsim10^5$ {\cmc} (see Fig.~\ref{fig:3}). As a result, isentropic instability occurs at $R_{\rm V}=5.5$. However, perhaps there are objects in the interstellar medium with $R_{\rm V}>5$ for $n_0<10^4$ {\cmc} or with $R_{\rm V}<4$ for $n_0>10^5$ {\cmc}.
 
\begin{figure*}
\begin{minipage}{1.3\columnwidth} 
\center{ \includegraphics[width=\columnwidth]{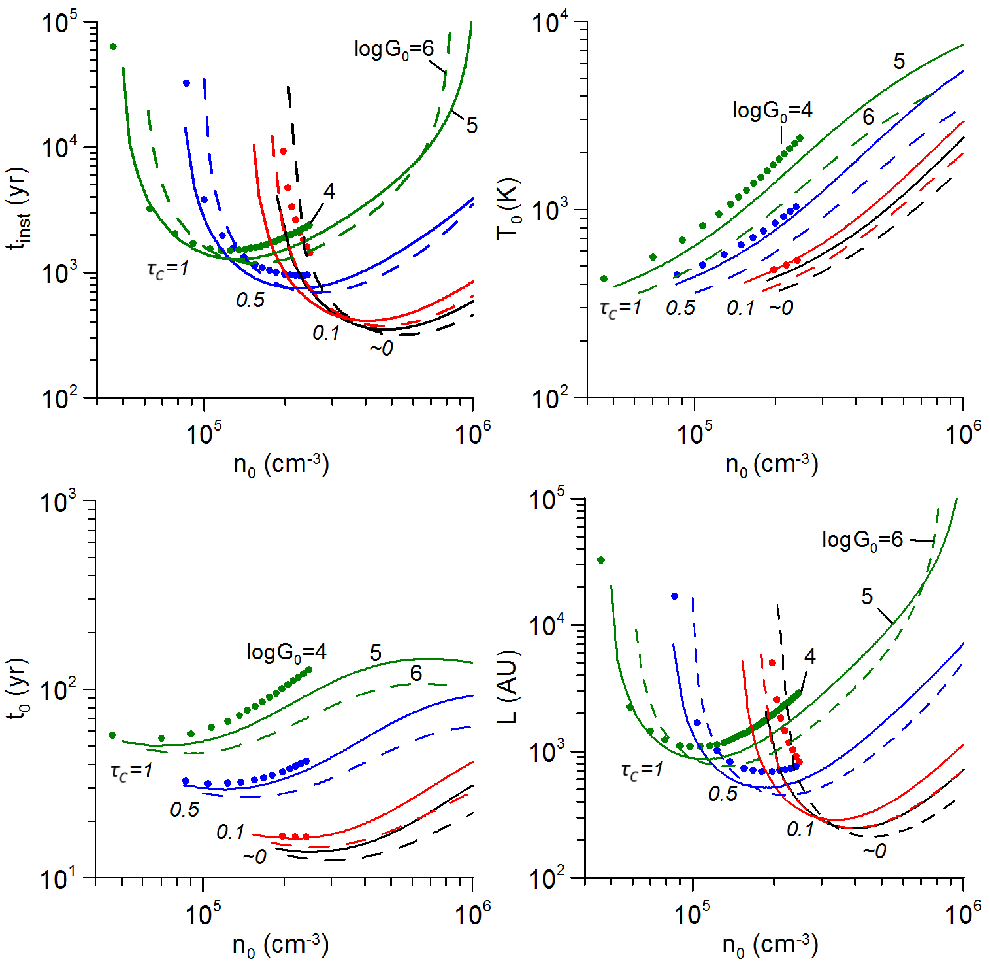} \\ (a)}
\end{minipage}
\begin{minipage}{0.7\columnwidth} 
\center{\includegraphics[width=\columnwidth]{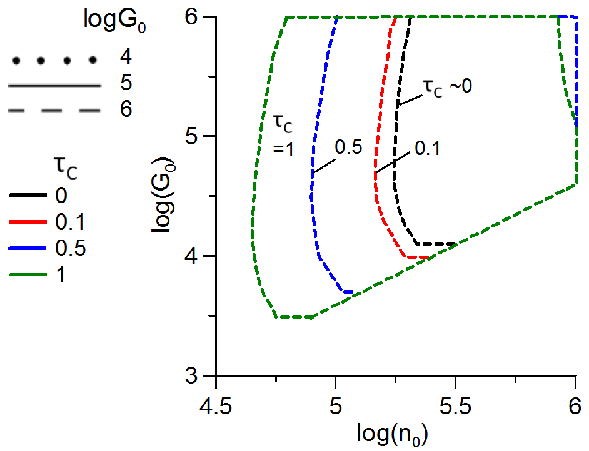}  \\ (b)}
   \caption{Functions  (a) $t_{\rm inst}$,  $T_0$, $t_0$, $L$ and (b) $G_0$ for the optical depth $0<\tau_{\rm C}\leqslant 1$ at $R_{\rm V}=5.5$,  $b_{\rm C}=3\times10^{-5}$, $\xi_{\rm C}=1.4\times10^{-4}$ and $\xi_{\rm O}=3.2\times10^{-4}$. The logarithmic relation between $G_0$ and $n_0$ in (b) shows the boundaries of the parameters causing instability (the boundaries correspond to the case $1/t_{\rm inst} \rightarrow 0$ and conditions \ref{eq:3} and  $G_0/n_0>0.04$ cm$^3$).}
    \label{fig:4}
\end{minipage}
  \end{figure*}

In the case of $R_{\rm V}=5.5$, instability criterion \ref{eq:2} is satisfied, when there are high intensities of the FUV fields $1.3\times10^4<G_0<10^6$ and high gas densities  $1.8\times10^5<n_0<10^6$ {\cmc} at temperatures $3.7\times10^2 <T_0 <2.5\times10^3$ K. More detailed distributions of $1/t_{\rm inst}$, 
$T_0$ and $G_0$ depending on $n_0$ are shown in Fig.  \ref{fig:3}. We obtain the following intervals:  characteristic perturbation growth time $3.1\times10^2<t_{\rm inst}<10^5$ yr (here an average value of the upper limit is given, although theoretically one can have $t_{\rm inst} \rightarrow  \infty$),  cooling time $12<t_0 <34$ yr and distance covering the locations of primary and secondary waves $2.1\times10^2<L<10^5$ au (also length-scale $4<l_0<32$ au and $24<L/l_0<10^4$). These parameters are shown in Fig.~\ref{fig:4}
  for  $\tau_{\rm C} \rightarrow  0$ ($\beta \sim 0.5$).

Thus, we find that acoustic thermal instability in the surface layers of PDRs can occur when the gas density and intensity of the incident FUV field are high.  

Next, we explore how the range of parameters causing instability changes if the opacity of the cooling lines and variations of element abundances are considered.

\subsubsection{Opacity of the fine-structure lines}
\label{sec:opacity}

Strictly speaking, to consider  the opacity  effect consistently we should use the distribution of gas parameters in the atomic zone.   As  we know, the value $\tau_{12}$ depends on  the depth $z$ of the plane-parallel layer  (where $0<z<Z$, i.e. $z$ varies  from the ionization (I) front to the dissociation (D) front), the level populations of the coolant element  (which can be expressed through the density in all levels $n$) and the temperature $T$  \citep{TielensHollenbach1985}. The approximate structure of the \ion{H}{i} zone (the thickness $Z$, distributions of $n(z)$ and $T(z)$) are calculated by solving the problem of I-D front propagation  depending on the incident FUV field, dust properties and abundances of elements. This is a complex problem even for one particular object with one set of  parameters. For the purposes of our study, we need to consider a very wide range of PDRs, for which the structures of the atomic zones will be distinguished substantially from each other.
Therefore, we would like to simplify the estimate of the optical depth and not to produce the calculation of the  \ion{H}{i} zone structure. We assume that, for any combination of $n$ and $T$, there exists a z  position for which the optical depth $\tau_{12}$ takes any values in a given interval (known from the studies of PDRs: \citealt{Tielens2005}).   Presumably, $\tau_{12}$ can   successively take all interval values independently of $n$ and $T$.  This allows us  to consider the optical depth   as a parameter of the cooling function, with values within the allowable range. We suppose that this approach is acceptable as the first approximation for a wide objects variation.

The infrared fine-structure [\ion{C}{ii}] 158, [\ion{O}{i}] 63 and [\ion{O}{i}]  146 {\micron} lines in the atomic zone of PDRs are characterized by optical depths $\tau_{21}$ in the range 0--1 \citep{TielensHollenbach1985, Tielens2005}. When  $\tau_{21}$ increases, the escape probability $\beta(\tau_{21})$ decreases. Consequently, the total cooling  $\Lambda$ weakens and the heat-loss function $Q=\Gamma - \Lambda$ increases. As a result, the steady-state temperature $T_0$ rises when the density is constant \citep{TielensHollenbach1985}. This temperature behaviour can be seen in Fig.~\ref{fig:4}, where changes of all optical depths are expressed through variations $\tau_{\rm C}$, the depth of the [\ion{C}{ii}] 158 {\micron}  line.

Fig.~\ref{fig:4} demonstrates that the inclusion of opacity in the cooling lines expands the range of PDR parameters causing instability. When  $\tau_{\rm C}$ increases,  criterion \ref{eq:2} is satisfied for a large number of values  $G_0, n_0$ and $T_0$ and higher values of time $t_{\rm inst}$ and $t_0$. As a result, the largest depth $\tau_{\rm C}=1$ corresponds to the largest intervals of values $G_0, n_0$ and $T_0$. The lower and upper bounds of  $t_{\rm inst}$, $t_0$ and $L$ correspond to the values  $\tau_{\rm C} \rightarrow 0$ and $\tau_{\rm C}=1$, respectively.

Within the interval $0<\tau_{21}\lesssim1$, we find the minimum and maximum values of the parameters causing instability. Thus, we obtain the total ranges: densities $4.5\times10^4<n_0<10^6$ {\cmc}, FUV fields $3\times10^3<G_0<10^6$ (we select cases for $G_0/n_0>0.04$ cm$^3$) and temperatures $360 <T_0 <10^4$ K. We also obtain  the time intervals  $3.1\times10^2<t_{\rm inst}<10^6$ yr, $12<t_0<2\times10^2$ yr and length-scales $2.1\times10^2<L<10^6$ au ($4<l_0<3.4\times10^2$ au, $23<L/l_0<10^4$).

Consequently, the previous result  (Section \ref{sec:Rv}), where isentropic instability occurs in  dense PDRs and for high intensity of radiation field, is preserved (but the lower bounds
of values $n_0$, $G_0$ and $T_0$ decrease slightly).

\subsubsection{Carbon and oxygen abundances} 

\begin{figure*}
\begin{minipage}{2\columnwidth}
\center{ \includegraphics[width=\columnwidth]{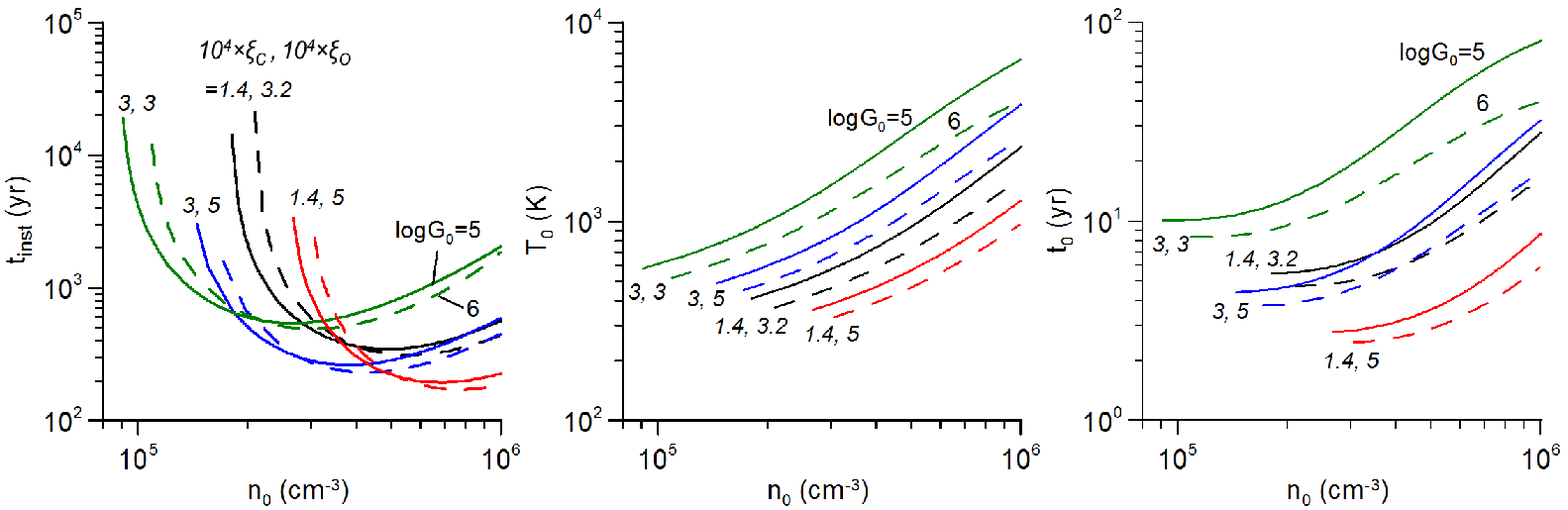} \\ (a)}
\end{minipage}
\vfill 
\begin{minipage}{0.72\columnwidth}
\center{\includegraphics[width=\columnwidth]{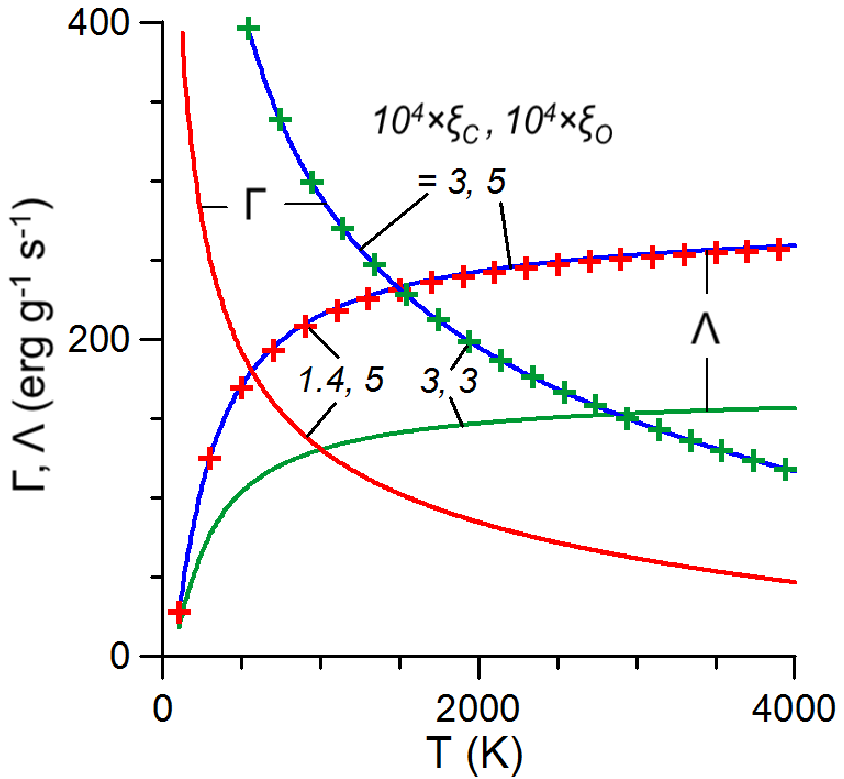} \\ (b)}
  \end{minipage}
  \begin{minipage}{0.72\columnwidth}
\center{\includegraphics[width=\columnwidth]{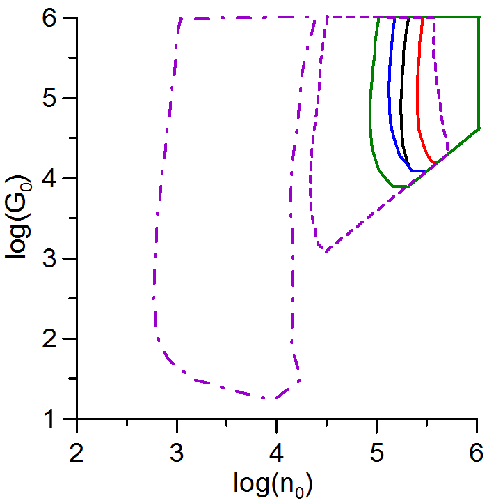} \\ (c)}
  \end{minipage}
    \begin{minipage}{0.55\columnwidth}
\center{\includegraphics[width=\columnwidth]{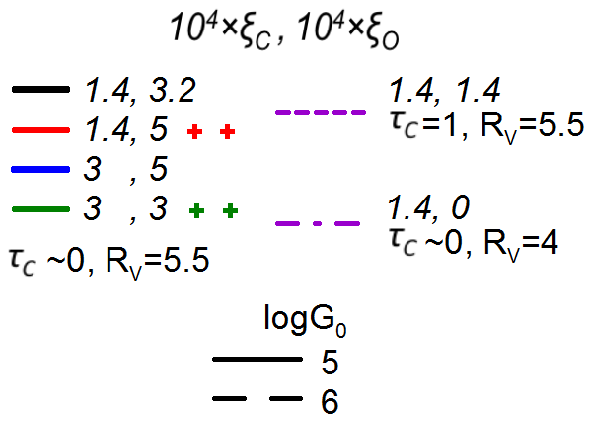}}
\end{minipage}
\caption{Influence of carbon and oxygen abundances  ($R_{\rm V}=5.5$,  $b_{\rm C}=3\times10^{-5}$, $\tau_{\rm C}\sim0$):  (a) $t_{\rm inst}$, $T_0$ and $t_0$ for $G_0=10^5$ and $10^6$  (solid and dashed curves); (b) heating $\Gamma$ and cooling $\Lambda$ rates for $G_0=10^5$, $n_0=5\times10^5$ {\cmc}.  Panel (c) shows the boundaries of parameters causing instability for $\xi_{\rm C}$, $\xi_{\rm O}$ as in panels (a) and (b) (solid curve) and for  case $\tau_{\rm C}=1$ at $\xi_{\rm C}, \xi_{\rm O} (\times10^{4})$ equal to 1.4, 1.4 (little dashed curve) and when cooling in the  [\ion{O}{i}] lines is neglected ($R_{\rm V}=4$, $b_{\rm C}=4\times10^{-5}$ and $\tau_{\rm C}\sim0$: dash-dotted line).}
    \label{fig:5}  
  \end{figure*}

The C and O abundances of PDRs have typical values $\xi_{\rm C}=1.4\times10^{-4}$ and $\xi_{\rm O}=3.2\times10^{-4}$\citep{Cardelli1996, Meyer1998}. To find the influence of $\xi_{\rm C}$ and $\xi_{\rm O}$ on the parameters causing instability, we consider variations of the abundances within the ranges used in early studies of PDRs, i.e. $\xi_{\rm C}=(1.4$--$3)\times10^{-4}$ and $\xi_{\rm O}=(3$--$5)\times10^{-4}$ \citep{TielensHollenbach1985, Wolfire1995}.  The main results are shown in Fig.~\ref{fig:5}.
 
The carbon abundance influences  gas cooling and heating (where  $\xi_{\rm C}$ governs the electron density $n_e$). However, in a medium with high density ($n\gtrsim10^5$ {\cmc}), the cooling in the [\ion{O}{i}] 63 {\micron} line is significantly larger than that in the  [\ion{C}{ii}] 158 and [\ion{O}{i}]  146 {\micron} lines \citep{TielensHollenbach1985, Burton1990}. Therefore, the contribution of carbon to the total cooling $\Lambda$ is very small. The dependence of photoelectron emission on electron density $n_e$ is well known, i.e. a decrease of $n_e$ leads to a decrease in total heating $\Gamma$. As a result, $\xi_{\rm C}$ reduction causes a decrease of the steady-state temperature $T_0$ obtained from the equation $\Gamma-\Lambda=0$. This property is shown in Fig.~\ref{fig:5}(b),  which presents a comparison of curves $\xi_{\rm C}$, $\xi_{\rm O}$ ($\times10^4$) between values 3, 5 and 1.4, 5. 
 At the same time, the oxygen only influences the gas cooling. Therefore, $\xi_{\rm O}$ reduction leads to a decrease in heating $\Gamma$ and hence leads to an increase in $T_0$ (see Fig.~\ref{fig:5}(b) when $\xi_{\rm C}$, $\xi_{\rm O}$ ($\times10^4$) are equal to 3,  5 and 3, 3). We note that the influence of a general decrease of  C and O abundances on $T_0$ is established by direct calculations of  the $\Gamma$ and $\Lambda$ functions.
 
The variations of $\xi_{\rm C}$ and $\xi_{\rm O}$   change the range of parameters causing instability (see Fig.~\ref{fig:5}). However, even if we take into account the opacity of cooling lines, then the orders of  the values  $n_0, G_0, T_0$, $t_{\rm inst}, t_0$, and $L$ are comparable with the corresponding orders for typical abundances $\xi_{\rm C}$ and $\xi_{\rm O}$ (see Section \ref{sec:opacity}). Thus,  for  $\xi_{\rm C}=(1.4$--$3)\times10^{-4}$ and $\xi_{\rm O}=(3$--$5)\times10^{-4}$ at $0<\tau_{21}\lesssim1$, we obtain the following  total intervals:  densities $2.2\times10^4<n_0<10^6$ {\cmc}, FUV fields $1.3\times10^3<G_0<10^6$ (when $G_0/n_0>0.04$ cm$^3$) and temperatures $322<T_0<10^4$ K.  We also obtain the characteristic perturbation growth time $1.7\times10^2<t_{\rm inst}<10^6$ yr, cooling time $7<t_0<4.5\times10^2$ yr and distance covering the locations of primary and secondary waves $10^2<L<10^6$ au ($3<l_0<7.5\times10^2$ au, $23<L/l_0<10^4$).  
 
The greatest change in the range of parameters causing instability is induced by a significant reduction of the oxygen abundance. We considered the limiting situation, when the fine-structure [\ion{O}{i}] 63 and [\ion{O}{i}]  146 {\micron}  lines are neglected completely (see Fig.~\ref{fig:5}(c)  for $\xi_{\rm C}$, $\xi_{\rm O}$ ($\times10^4$) are equal to 1.4, 0). In this case, the isentropic instability criterion is satisfied for intermediate densities $6\times10^2<n_0<2.5\times10^4$ {\cmc} and for a wide range of FUV fields $20<G_0<10^6$ at temperatures $1.1\times10^2<T_0<9\times10^3$ K. Nevertheless, the thermal balance model of the \ion{H}{i} zone in PDRs, in which the oxygen fine-structure lines were ignored at intermediate density  ($n>10^2$ {\cmc}), requires theoretical and observational arguments. We could neglect the [\ion{O}{i}] 63  {\micron} emission compared with the [\ion{C}{ii}] 158  {\micron} line only for low-density PDRs, i.e. diffuse gas with $n<10^2$ {\cmc}  \citep{Hollenbach1991}. However, for such low densities the isentropic instability criterion \ref{eq:2} is not satisfied. 
Diffuse clouds usually have another model of  chemical and energy balance \citep{Wolfire1995, Wolfire2003}, which differs from the case of dense clouds.  Moreover, thermal instability may also occur in diffuse gas, but in another mode, the isobaric instability \ref{eq:1}. 

\subsubsection{General results} 

We found the conditions for which the isentropic instability criterion \ref{eq:2} on the surface layer of a PDR is satisfied. We used a model of the energy balance with photoelectric heating from interstellar grains and cooling through the fine-structure [\ion{C}{ii}] 158, [\ion{O}{i}] 63 and [\ion{O}{i}]  146 {\micron}  lines. For a wide range of parameters, which characterize the generalized heat-loss function $Q = \Gamma-\Lambda$, we obtained the following results.
 
 \begin{itemize}
 \item  Isentropic  thermal instability can occur if the gas density and intensity of the incident FUV field are high. We estimated ranges of the FUV field, density, and temperature when the opacity of the cooling lines ($0<\tau_{21}\lesssim1$ ) is taken into account and C and O abundances are typical: $\xi_{\rm C}=1.4\times10^{-4}$ and $\xi_{\rm O}=3.2\times10^{-4}$. These intervals are
  \begin{equation}
  \begin{split}
& 3\times10^3<G_0<10^6 \, , 4.5\times10^4<n_0<10^6 \textrm{{\cmc}} \, , \\
&  360<T<10^4 \, \textrm{K} \,. 
 	\label{eq:4}
\end{split}	
\end{equation}
We also obtained ranges of characteristic perturbation growth time $3.1\times10^2<t_{\rm inst}<10^6$ yr,  cooling time $12<t_0<2\times10^2$ yr and distance that characterizes  secondary wave formation $2.1\times10^2<L<10^6$ au (for initial perturbation wavelength $4<\lambda<3.4\times10^2$ au, where $\lambda=l_0$).  

 \item Variations of carbon and oxygen abundances $\xi_{\rm C}=(1.4$--$3)\times10^{-4}$, $\xi_{\rm O}=(3$--$5)\times10^{-4}$  slightly change the ranges of the parameters causing instability, but the ranges correspond to within the order of their values in the case of typical abundances ($\xi_{\rm C}=1.4\times10^{-4}$ and $\xi_{\rm O}=3.2\times10^{-4}$). If we take into account the opacity of the cooling lines, then we obtain the intervals 
  \begin{equation}
  \begin{split}
&1.3\times10^3<G_0<10^6 \, , 2.2\times10^4<n_0<10^6 \textrm{{\cmc}} \, , \\
&  322<T<10^4 \, \textrm{K} \,. 
 	\label{eq:5}
\end{split}	
\end{equation}  
  \item  A significant decrease of the oxygen contribution to gas cooling gives the greatest impact on the change of the parameters causing isentropic instability.
  \end{itemize}

 \section{Examples of observed PDR{\small s} where instability can occur} 
 \label{sec:4}
    
{The assumption that the turbulent motion in an atomic interstellar medium can be caused by thermal instability was discussed earlier by \citet{KritsukNorman2002, Brandenburg2007, IwasakiInutsuka2014}. These articles studied the isobaric mode of thermal instability and considered the heat-loss rate $Q$ for a diffuse atomic gas \citep{Wolfire1995}. However, as we shall see below, turbulent motions in a dense PDR can  also  be caused by the isentropic type of instability.}
  
 The results obtained in the previous sections can be used to find out whether instability of travelling waves arises in some observed PDRs. Let us consider examples of these PDRs and discuss the corresponding estimates of  the main parameters causing instability. The main parameters are the FUV field $G_0$, steady-state density of  the atomic gas $n_0$ and abundances $\xi_{\rm C}$ and  $\xi_{\rm O}$. The gas temperature $T_0$ is determined from the equation of  energy balance and depends on the optical depths of cooling lines.  The observed PDRs with parameters satisfying the ranges \ref{eq:4} and \ref{eq:5} are given in Table~\ref{tab:2}. 
   \begin{table*}
\caption{Examples of the observed PDRs.}
  \label{tab:2}
  \begin{tabular}{ccccccccc}
    \hline
    Object & PDR& $G_0$ & n & T & $\xi_{\rm C}$ & $\xi_{\rm O}$ & R & D \\
       &   &     & {\cmc} & K & $\times10^4$ & $\times10^4$ & pc & pc  \\
    \hline
    1 & Orion Bar& [1-4](4) & [0.5-1](5) & [0.5-1](3) & 3 & 5, 4$^{a}$ & 0.02$^{b}$ & 0.3\\
    2 & NGC 2023 S & [3-6](3) & [0.5-2](5) & [0.3-1](3) & 1.4 & 3.2 & 0.004$^{c}$ & 0.04\\
    3 & NGC 7023 NW & [2.6-7.7](3) & [0.5-2](5) & [3-5](2) & 1.6 & 3.2 & 0.02$^{d}$ & 0.1\\
    4 & Mon R2 & [0.5-1](5) & [0.4-4](5) & [3-6](2) & 1.6  & 3.2& 0.001$^e$ & -\\
    5 & Carina N$^f$  & [0.7-1.6](4) & [2-10](5) & [3-6](2) & 1.6  & 3.2 & - & - \\
    \hline
  \end{tabular}  
  \begin{tabular}{l}
  {\it Notes}. \\
Numbers in parentheses: [1-4](4) corresponds to the interval $10^4$ -- $4\times10^4$. \\
 The last two columns are approximate sizes of  the PDR atomic layers, where R and D are sizes in the radial and perpendicular directions. \\
 \textbf{References}. Objects: \\
 \textbf{1}. \citealt{Tauber1994, YoungOwl2000}; $^a$ \citealt{Pellegrini2009}; $^b$ \citealt{Bernard-Salas2012}. \\
 \textbf{2}. $^c$ \citealt{Sheffer2011, Sandell2015}. \textbf{3}. \citealt{Joblin2010}; $^d$ \citealt{Pilleri2012, Okada2013}.\\
  \textbf{4}. \citealt{Berne2009}; $^e$ \citealt{Pilleri2014, Okada2013}.  \textbf{5}. \citealt{Brooks2003, Kramer2008};  \\
 $^f$ according to  \citealt{Okada2013}  we assume the absence of the [\ion{O}{i}] 146 {\micron} emission and  $b_{\rm C}=0$ at $R_{\rm V}=5.5$.  
  \end{tabular} 
  \end{table*}

   \begin{figure*}
\begin{minipage}{2\columnwidth}
\center{ \includegraphics[width=\columnwidth]{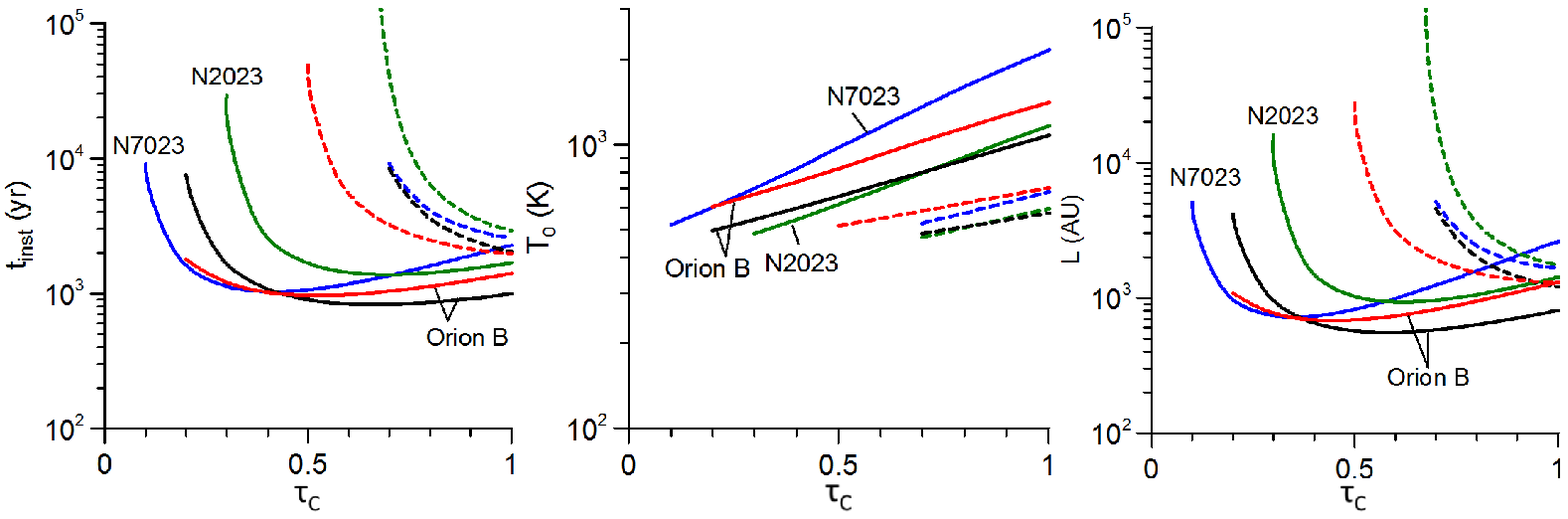} \\ (a)}
\end{minipage}
  \hfill
  \begin{minipage}{2\columnwidth}
\center{\includegraphics[width=\columnwidth]{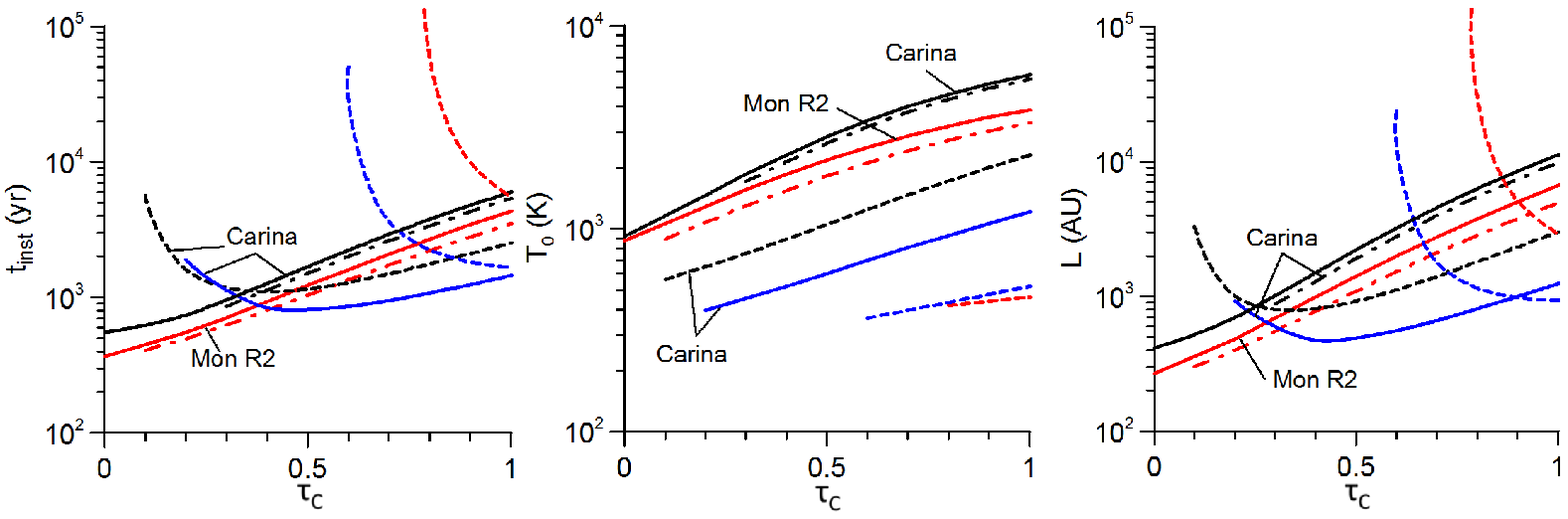} \\ (b)}
  \end{minipage}
    \hfill
    \begin{minipage}{2\columnwidth}
\center{\includegraphics[width=0.9\columnwidth]{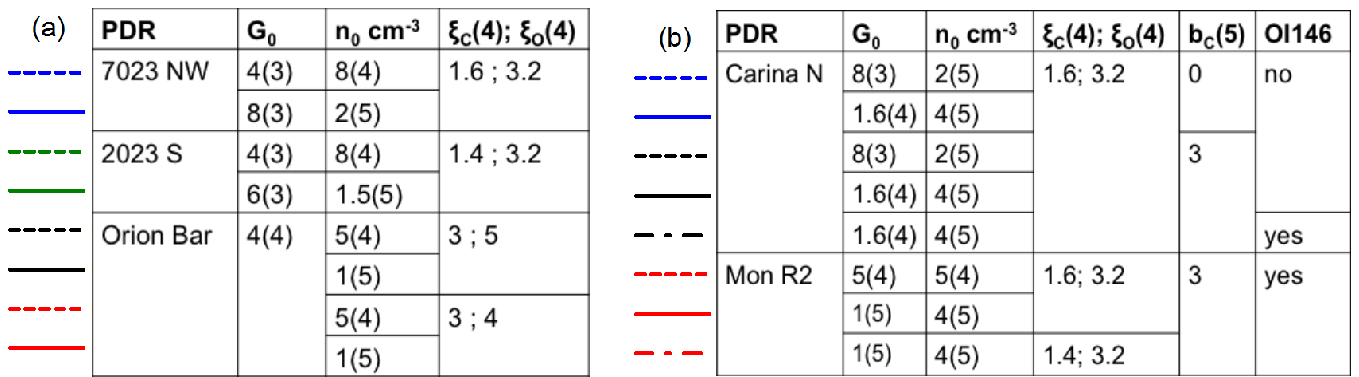}}
\end{minipage}
\caption{Functions $t_{\rm inst}$, $T_0$, and $L$ of $\tau_{\rm C}$ on surfaces of PDRs with parameters similar to values from Table~\ref{tab:2}:  (a) NGC 7023 NW, NGC 2023 S and Orion Bar  for $R_{\rm V}=5.5$, $b_{\rm C}=3\times10^{-5}$;  (b)  Carina N and Mon R2 for $R_{\rm V}=5.5$. Note numbers in parentheses: 4(3)=$4\times10^{3}$.}
    \label{fig:6}  
  \end{figure*}

 Fig.~\ref{fig:6} for each of the PDRs shows  functions $T_0$, $t_{\rm inst}$ and $L$  for which criterion \ref{eq:2} is satisfied.  We found typical values of the gas temperature $T_0\sim3\times10^2$--$2\times10^3$ K, characteristic perturbation growth time $t_{\rm inst}\sim10^3$ --$10^4$ yr and distance characterizing secondary wave appearance $L\sim2\times10^2$--$10^4$ au $=10^{-3}$ -- $5\times10^{-2}$  pc {at the wavelength $\lambda \sim 6\times10^{-5} - 2\times10^{-3}$  pc}.  We see that the average scale $L$ is less than (or the same order as)  the atomic layer sizes $R$ or $D$ in Table~\ref{tab:2}. Since the amplitude of waves for propagation time $t\sim t_{\rm inst}$  is close to the amplitude of the saturation mode \citep{Krasnobaev2016}, we can expect a significant influence of autowaves on the velocity dispersion if $R\gtrsim L$ or $D \gtrsim L$ and $t\gtrsim t_{\rm inst}$. Next we consider the influence of isentropic instability on the velocity field \textbf{v}, density $\rho$ and temperature $T$ in detail. 
  
As was shown in Section \ref{sec:unstable}, acoustic  instability is characterized by presence of multiple shock waves in the gas. The corresponding relative variations  ${ u/u_0}$, $\rho/\rho_0$ and $T/T_0$  behind the shocks have amplitudes in the range 0.1--0.5, where the maximum of the values  corresponds to the saturation amplitude \citep{Krasnobaev2016}. Consequently, the turbulent velocity $u_{\rm turb}$, which has the same order as the gas velocity behind the shock wave,  is approximately  equal to several kilometers per second (see below for details).
Due to collisions of the shock waves with sharp  boundaries, such as ionization and dissociation fronts, the value of the  turbulent velocity $u_{\rm turb}$ can be higher \citep{Chernyi1988}. 
These velocity variations are quite accessible to  observations \citep{MieschBally1994, Yoshida2010}. Multiple shock waves can be observed morphologically as filamentary or reticulate structures, not only in an  \ion{H}{i} zone but also in ionized gas (due to the penetration of perturbations into an  \ion{H}{ii} region).  If acoustic  instability occurs, then the density and temperature in filamentary structures is higher than that in the surrounding gas. Such structures are observed, for example, in RCW 120 \citep{Zavagno2007, Deharveng2009}. They could  be formed for a time shorter than the age of RCW 120. 
We take into account the fact that the density and temperature distributions in RCW 120 are sufficiently inhomogeneous. Using RCW 120 estimates from the literature  \citep{Zavagno2007,  Torii2015}, we find that in dense clouds we have $n_0\sim10^5$ {\cmc}, $T_0\sim550$ K and in a less dense medium we have $n_0\sim10^4$ {\cmc}, $T_0\sim140$ K. According to the PDR model of RCW 120 by \citet{Rodon2015}, we have the density $n_0\sim2\times10^4$ {\cmc} and FUV flux $G_0\sim6\times10^2$. RCW 120 parameters differ insignificant from  the parameters causing isentropic instability (see ranges \ref{eq:4} and \ref{eq:5}).  For example, if we assume $n_0=7\times10^4$ {\cmc} and $G_0=3\times10^3$  then,   using our energy balance model (Section \ref{sec:2}) and criterion  \ref{eq:2}, we find $T_0=5.2\times10^2$ K, $t_{\rm inst}\sim7\times10^3$ yr,   $L\sim4\times10^3$ au $=2\times10^{-2}$ pc for  $\tau_{\rm C}\sim1$ and $\xi_{\rm C}=1.4\times10^{-4}$ and $\xi_{\rm O}=3.2\times10^{-4}$. The characteristic perturbation growth time $t_{\rm inst}$ is less than the estimated age of the  \ion{H}{ii}  region, which is greater than $4\times10^5$ yr, and the length-scale $L$ is less than the thickness of the surface layer $R\sim5\times10^{-2}$ pc \citep{Zavagno2007, Torii2015}.   

 The presence of multiple shocks  (autowaves) can also  be manifested as significant changes of gas parameters (density, velocity and temperature) on very small spatial scales that are the same order as the thicknesses of the corresponding shock fronts $d_{\rm S} \sim 10^{15}/n$ cm \citep{Landau1987}, where  $d_{\rm S} < 10^{13}$ cm $\sim 3 \times 10^{-6}$ pc at $n>10^2$ \cmc. 
The existence of similar fluctuations is shown by the  analysis of turbulent velocities in the Orion Nebula \citep{Ferland2012}.  Observation of this object in the atomic zone of the PDR gives $u^{\rm Orion}_{\rm turb} \approx 5$ \kms  \, at $T\approx10^3$ K. For such a gas temperature, the adiabatic sound speed is $a^{\rm Orion}_0\approx 3.7$ \kms  (the mean mass per particle is equal to $1.3$). Since the turbulent velocity $u_{\rm turb}$ has the same order as the gas velocity $u$ (moreover, it can be estimated as $2  u$, \citealt{Ferland2012}), its magnitude corresponds  to $u_{\rm turb} \sim 2  u \sim 2 \times0.5 u_0=0.8 a_0$. Therefore, in the case of the possible growth of isentropic perturbations, we can obtain the turbulent velocity in this PDR as  $u_{\rm turb} \sim 3$ \kms.  Though  the estimate $u_{\rm turb}$ is  slightly less than the observed velocity $u^{\rm Orion}_{\rm turb}$, we have satisfactory  conformity in these values.

The study of the observation data in this section was obtained under the assumption that we can neglect thermal conductivity. This assumption is valid if  $t_{\rm inst} \ll t_{\rm h}$, where the conductive time is $t_{\rm h}=\lambda^2 n k_ {\rm B} /(\gamma-1)\kappa$  and the coefficient of thermal conductivity for atomic gas \citep{Lang1974} is $\kappa={5 k_ {\rm B}}/{2 m_{\rm H}} \, 5.7 \times 10^{-5} \,\sqrt{T}$ (ergs s$^{-1}$ K$^{-1}$ cm$^{-1}$).   For  typical PDR parameters such as  $n \sim 10^5$ cm$^{-3}$ and $T \sim 10^3$ K at the average time $t_{\rm inst}\sim10^3$ yr (for example the Orion Bar: Table~\ref{tab:2}, Fig.~\ref{fig:6}), we find   that $t_{\rm inst}<t_{\rm h}$ is satisfied when wavelength $\lambda>\lambda_{\rm cr}=10^{-6}$ pc (where $\lambda_{\rm cr}^2 n k_ {\rm B} /(\gamma-1)\kappa=t_{\rm inst}$). Since, for the PDRs studied above,  we have the wavelength of adiabatic perturbations $\lambda \sim 6\times10^{-5}-2 \times10^{-3}$  pc,  for such conditions the influence of thermal conductivity is insignificant. Notice that the  critical wavelength $\lambda_{\rm cr}$ is similar to the length from Field's theory  \citep{Field1965}, i.e. $\lambda_{\rm F}=2\pi/\sqrt{{\rho_0}(Q_T+{\rho_0 Q_\rho}/{(\gamma-1)T} )/{\kappa}}$. On the other hand, in a dense PDR, perturbations with a very small scale of the order of the shock-front thickness  $d_{\rm S}$  will be damped under the influence of conductivity.

We emphasize some limitations and uncertainties that appear in the development of our model. Consistent treatment  of the opacity effect assumes that there are distributions of $n (z)$ and $T (z)$ in the atomic zone that correspond to one set of values $\tau_{12} (z, n, T)$ for the cooling lines. The resulting values obtained by this approach ($n_0$, $T_0$ and $\tau_{12}$ in all the atomic zone ) are contained among the values found in our rough approximation (see Section \ref{sec:opacity}). In other words, by a consistent  treatment  we can obtain a smaller number of resulting  values  (up to a total absence) satisfying the instability criteria compared with the case of our approximation (Fig.~\ref{fig:6}). The results of the rough approximation give a larger number of combinations of parameters that characterizes the medium in a state of thermal instability than is the case for real PDRs. However, our approach allows us to estimate the order and the approximate values of these parameters.

Another significant limitation is the neglect of large-scale motions in PDRs. If we take these motions into account,  then the energy balance and consequently the gas temperature and density can  change.  We cannot exclude completely the influence of  the magnetic field, radiation pressure and cosmic rays \citep{Pellegrini2009} on the growth and structure of perturbations. However, detailed information about these processes is currently unavailable for most of PDRs.

\section{Conclusions}

The general aim of this work was to determine the implementability of isentropic thermal instability in the atomic surface layers of PDRs. Our research has verified it.
 
 \begin{itemize}
 \item We proposed a model of energy balance on the surface of a PDR, in which  gas is heated by photoelectron emission from dust grains and  cooled through the fine-structure excitation of ions and atoms by atomic hydrogen impact. We have taken into account the intensity of the far-ultraviolet radiation penetrating to the PDR, the optical depth of fine-structure lines and  variations in abundances of heavy elements.
 \item We found that, for typical abundances of elements, the medium will be thermally unstable for a dense PDR ($n_0>2\times10^4$ {\cmc}) and high intensity of the far-ultraviolet field ($G_0>10^3$).  When we take into consideration the opacity of the cooling lines, the intervals of key parameters ($G_0, n_0$ and $T_0$) causing instability are expanded.  We also found that the instability criterion depends significantly  on the relations of carbon and oxygen abundances.
 \item We gave examples of observed dense PDRs  that are affected by high-intensity FUV flux  and in which isentropic instability can occur. We found the characteristic perturbation growth time $t_{\rm inst}\sim10^3$--$10^4$ yr  and distance covering the locations of primary and secondary waves $L\sim10^{-3}$ -- $5\times10^{-2}$ pc.  For objects older than $t_{\rm inst}$ and with the scale of the atomic zone greater than $L$, we described the features of the instability (for example, RCW 120). These features include the presence of multiple shock waves and filamentous structures with higher density and temperature than the surrounding medium. 
 \end{itemize}





\bibliographystyle{mnras}
\bibliography{example} 

\begin{thebibliography}{}
\bibitem[\protect\citeauthoryear{Bakes \& Tielens}{1994}]{BakesTielens1994} Bakes E.~L.~O., Tielens A.~G.~G.~M., 1994, \apj, 427, 822
\bibitem[\protect\citeauthoryear{Baranov \& Krasnobaev}{1977}]{BarKrasn1977} Baranov V.~B., Krasnobaev K.~V., 1977, Hydrodynamic Theory of Cosmic Plasma. Nauka, Moscow, p. 335 [in Russian] 
\bibitem[\protect\citeauthoryear{Bernard-Salas et al.}{2012}]{Bernard-Salas2012} Bernard-Salas J. et al.,  2012, \aap, 538, A37
\bibitem[\protect\citeauthoryear{Berne et al.}{2009}]{Berne2009} Berne O.,  Fuente A.,  Goicoechea J.~R., Pilleri P., Gonzalez-Garcia M.,   Joblin C.,  2009, \apj, 706, L160
\bibitem[\protect\citeauthoryear{ Brandenburg, Korpi \& Mee}{2007}]{Brandenburg2007}  Brandenburg A., Korpi M.~J., Mee A., 2007, \apj, 654, 945
\bibitem[\protect\citeauthoryear{Brooks et al.}{2003}]{Brooks2003} Brooks K.~J., Cox P., Schneider N., Storey J.~W.~V.,  Poglitsch A., Geis N.,  Bronfman L., 2003, \aap, 412, 751
\bibitem[\protect\citeauthoryear{Burton, Hollenbach \& Tielens}  
{1990}]{Burton1990} Burton M.~G., Hollenbach D.~J., Tielens A.~G.~G.~M., 1990, \apj, 365, 620
\bibitem[\protect\citeauthoryear{Cardelli, Clayton \& Mathis} 
{1989}]{CardelliClaytonMathis1989}  Cardelli J.~A., Clayton G.~C., Mathis J.~S. 1989, \apj, 345, 245
\bibitem[\protect\citeauthoryear{Cardelli et al.}{1996}]{Cardelli1996} Cardelli J.~A., Meyer D.~M., Jura M., Savage B.~D., 1996, \apj, 467, 334
\bibitem[\protect\citeauthoryear{Chernyi}{1988}]{Chernyi1988} Chernyi G.~G., 1988, Gas Dynamic. Nauka, Moscow, p. 424 [in Russian]
\bibitem[\protect\citeauthoryear{de Jong, Dalgarno \& Boland} 
{1980}]{Jong1980} de Jong T.,  Dalgarno A., Boland W., 1980, \aap, 91, 68
\bibitem[\protect\citeauthoryear{Deharveng et al.}{2009}]{Deharveng2009} Deharveng L., Zavagno A., Schuller F., Caplan J., Pomares M., De Breuck C., 2009, \aap,  496, 177
\bibitem[\protect\citeauthoryear{Draine \& Bertoldi}{1996}]{DraineBertoldi1996} Draine B.~T., Bertoldi F., 1996, \apj, 468, 269
\bibitem[\protect\citeauthoryear{Ferland et al.}{2012}]{Ferland2012} Ferland G. J., Henney W. J., O'Dell C. R., Porter R. L., van Hoof  P. A. M., Williams R. J. R., 2012,  \apj, 757,79
\bibitem[\protect\citeauthoryear{Field}{1965}]{Field1965} Field G.~B., 1965, \apj, 142, 531
\bibitem[\protect\citeauthoryear{Field}{1969}]{Field1969} Field G.~B., Goldsmith D.W. \& Habing H.J., 1969, \apj, 155, L149
\bibitem[\protect\citeauthoryear{Habing}{1968}]{Habing1968} Habing H.~J., 1968, \bain, 19, 421
\bibitem[\protect\citeauthoryear{Hollenbach \& Tielens}{1999}]{HollenbachTielens1999} Hollenbach D.~J.,  Tielens A.~G.~G.~M., 1999, Rev. Mod. Phys., 71, 173 
\bibitem[\protect\citeauthoryear{Hollenbach, Takahashi \& Tielens}
{1991}]{Hollenbach1991} Hollenbach D.~J., Takahashi T., Tielens A.~G.~G.~M., 1991, \apj, 377, 192
\bibitem[\protect\citeauthoryear{Iwasaki \& Inutsuka}{2014}]{IwasakiInutsuka2014} Iwasaki K.,  Inutsuka S., 2014, \apj, 784, 115
\bibitem[\protect\citeauthoryear{Joblin et al.}{2010}]{Joblin2010} Joblin C. et al., 2010,  \aap, 521, L25
\bibitem[\protect\citeauthoryear{Kaplan \& Pikelner}{1979}]{KaplanPikelner1979} Kaplan S.~A.,  Pikelner S.~B., 1979, Physics of Interstellar Medium. Nauka, Moscow, p. 591 [in Russian]
\bibitem[\protect\citeauthoryear{Kramer et al.}{2008}]{Kramer2008} Kramer C. et al., 2008, \aap, 477, 547
\bibitem[\protect\citeauthoryear{Krasnobaev \& Tarev}{1987}]{KrasnobaevTarev1987} Krasnobaev K.~V., Tarev V.~Y., 1987, Astron. J., 64, 1210 [in Russian] 
\bibitem[\protect\citeauthoryear{Krasnobaev, Sysoev \& Tarev } 
{1994}]{Krasnobaev1994} Krasnobaev K.~V., Sysoev N.~E.,  Tarev V.~Yu., 1994, 
Nuclear Physics, Cosmic Radiation, Astronomy. Mosk. Gos. Univ., Moscow,  p. 222 [in Russian] 
\bibitem[\protect\citeauthoryear{Krasnobaev et al.}{2016}]{Krasnobaev2016} Krasnobaev  K.~V.,  Tagirova R.~R., Arafailov S.~I.,  Kotova G.~Yu., 2016, Astron. Lett., 42, 460
\bibitem[\protect\citeauthoryear{Kritsuk \& Norman}{2002}]{KritsukNorman2002}  Kritsuk A.~G., Norman M.~L., 2002, \apj, 569, L127
\bibitem[\protect\citeauthoryear{Landau  \&  Lifshitz}{1987}]{Landau1987} Landau L. D.,  Lifshitz E. M., 1987, Course of Theoretical Physics, Vol. 6: Fluid Mechanics, 2nd ed. Pergamon Press, Oxford
\bibitem[\protect\citeauthoryear{Lang K.R.}{1974}]{Lang1974} Lang K.R., 1974, Astrophysical Formulae. Springer-Verlag, Berlin
\bibitem[\protect\citeauthoryear{Li \& Draine}{2001}]{LiDraine2001} Li  A., Draine B.~T., 2001, \apj, 554, 778
\bibitem[\protect\citeauthoryear{Meyer, Jura \& Cardelli} 
{1998}]{Meyer1998} Meyer D.~M., Jura M., Cardelli J.~A., 1998, \apj, 493, 222
\bibitem[\protect\citeauthoryear{Miesch \& Bally}{1994}]{MieschBally1994} Miesch M.~S., Bally J., 1994, \apj, 429, 645
\bibitem[\protect\citeauthoryear{Molevich et al.}{2011}]{Molevich2011}  Molevich N.~E., Zavershinsky D.~I., Galimov R.~N., Makaryan V.~G.,  2011, \apss, 334, 35 
\bibitem[\protect\citeauthoryear{Nakariakov, Mendoza-Briceno \& Ibanez} 
{2000}]{Nakariakov2000} Nakariakov V.~M.,  Mendoza-Briceno C.~A., Ibanez M.~H., 2000, \apj, 528, 767
\bibitem[\protect\citeauthoryear{Okada, Pilleri \& Berne}
{2013}]{Okada2013} Okada Y.,   Pilleri P., Berne O., 2013, \aap,  553, A2 
\bibitem[\protect\citeauthoryear{Oppenheimer}{1977}]{Oppenheimer1977} Oppenheimer M., 1977,  \apj, 211, 400
\bibitem[\protect\citeauthoryear{Osterbrock \& Ferland}{2006}]{OsterbrockFerland2006} Osterbrock D.~E.,   Ferland G.~J., 2006, Astrophysics of gaseous nebulae and active galactic nuclei. University Science Books, Mill Valley, CA, p. 480
\bibitem[\protect\citeauthoryear{Parker}{1953}]{Parker1953} Parker E.~N., 1953, \apj, 117, 431
\bibitem[\protect\citeauthoryear{Pellegrini et al.}{2009}]{Pellegrini2009} Pellegrini E.~W.,  Baldwin J.~A.,  Ferland G.~J.,  Shaw G., Heathcote S., \apj, 693, 285
\bibitem[\protect\citeauthoryear{Pilleri et al.}{2012}]{Pilleri2012} Pilleri P.,  Montillaud J., Berne O.,  Joblin C., 2012, \aap, 542, A69
\bibitem[\protect\citeauthoryear{Pilleri et al.}{2014}]{Pilleri2014} Pilleri P. et al., 2014, \aap, 561, A69
\bibitem[\protect\citeauthoryear{Rodon et al.}{2015}]{Rodon2015} Rodon J.~A., Zavagno A., Baluteau J.~P., Habart E., Kohler M., Le Bourlot J., Le Petit  F.; 
Abergel A., 2015, \aap,  579, A10
\bibitem[\protect\citeauthoryear{Sandell et al.}{2015}]{Sandell2015} Sandell G., Mookerjea B., Gusten R., Requena-Torres M.~A., Riquelme D., Okada Y., 2015, \aap, 578, A41
\bibitem[\protect\citeauthoryear{Shaw et al.}{2009}]{Shaw2009} Shaw G., Ferland G.~J., Henney W.~J., Stancil P.~C., Abel N.~P.,
 Pellegrini E.~W., Baldwin J.~A., van Hoof P.~A.~M., 2009,  \apj, 701, 677
 \bibitem[\protect\citeauthoryear{Shchekinov}{1979}]{Shchekinov1979} Shchekinov, Yu.~A., 1979, \sovast, 15, 224
\bibitem[\protect\citeauthoryear{Sheffer et al.}{2011}]{Sheffer2011} Sheffer Y.,  Wolfire M.~G.,  Hollenbach D.~J., Kaufman M.~J.,  Cordier M.,  2011, \apj, 741, 45 
\bibitem[\protect\citeauthoryear{Sofia et al.}{2004}]{Sofia2004} Sofia U.~J., Lauroesch J.~T., Meyer D.~M., Cartledge S.~I.~B., 2004, \apj,  605, 272
\bibitem[\protect\citeauthoryear{Spitzer}{1948}]{Spitzer1948} Spitzer L.~Jr., 1948, \apj, 107, 6
\bibitem[\protect\citeauthoryear{Sternberg et al.}{2014}]{Sternberg2014} Sternberg A., Le Petit F., Roueff E., Le Bourlot J., 2014,  \apj, 790, 10
\bibitem[\protect\citeauthoryear{Tauber et al.}{1994}]{Tauber1994} Tauber J.~A., Tielens A.~G.~G.~M., Meixner M.,  Goldsmith P., 1994,  \apj, 422, 136
\bibitem[\protect\citeauthoryear{Tielens}{2005}]{Tielens2005} Tielens A.~G.~G.~M., 2005, The Physics and Chemistry of the Interstellar Medium. Cambridge Univ. Press, Cambridge, p. 206
\bibitem[\protect\citeauthoryear{Tielens \& Hollenbach}{1985}]{TielensHollenbach1985} Tielens A.~G.~G.~M., Hollenbach D.~J., 1985, \apj, 291, 722
\bibitem[\protect\citeauthoryear{Torii et al.}{2015}]{Torii2015} Torii K. et al., 2015,  \apj, 806, 7
\bibitem[\protect\citeauthoryear{Vazquez-Semadeni et al.}{2003}]{Vazquez2003} Vazquez-Semadeni E., Gazol A., Passot T., Sanchez-Salcedo J.   in Falgarone E., Passot. T., eds,  Lecture Notes in Physics, Vol. 614, Turbulence and Magnetic Fields in Astrophysics. Springer, Berlin, p. 213
\bibitem[\protect\citeauthoryear{Weingartner \& Draine}{2001a}]{WeingartnerDraine2001a}Weingartner J.~C., Draine B.~T., 2001, \apj, 548, 296
\bibitem[\protect\citeauthoryear{Weingartner \& Draine}{2001b}]{WeingartnerDraine2001b} Weingartner J.~C., Draine B.~T., 2001, \apjs, 134, 263
\bibitem[\protect\citeauthoryear{Wolfire et al.}{1995}]{Wolfire1995} Wolfire M.~G., Hollenbach D., McKee C.~F., Tielens A.~G.~G.~M., Bakes E.~L.~O.,1995, \apj,  443, 152
\bibitem[\protect\citeauthoryear{Wolfire et al.}{2003}]{Wolfire2003} Wolfire M.~G., McKee C.~F., Hollenbach D.~J., Tielens A.~G.~G.~M., 2003,  \apj, 587, 278
\bibitem[\protect\citeauthoryear{Yoshida et al.}{2010}]{Yoshida2010} Yoshida A., Kitamura Y., Shimajiri Y., Kawabe R.,  2010, \apj, 718, 1019
\bibitem[\protect\citeauthoryear{Young Owl et al.}{2000}]{YoungOwl2000} Young Owl R.~C., Meixner M.~M., Wolfire M., Tielens A.~G.~G.~M., Tauber J., 2000, \apj, 540, 886
\bibitem[\protect\citeauthoryear{Zanstra}{1955}]{Zanstra1955} Zanstra H., 1955, in Vistas in Astronomy Vol. 1. Pergamon, New York, p. 256 
\bibitem[\protect\citeauthoryear{Zavagno et al.}{2007}]{Zavagno2007} Zavagno A., Pomares M., Deharveng L., Hosokawa T., Russeil D.,  Caplan J., 2007, \aap, 472, 835 
\end{thebibliography}





\bsp	
\label{lastpage}
\end{document}